\pdfoutput=1

\documentclass[11pt,twoside,a4paper,cmspaper,final,collab]{cms-tdr}
\def\svnVersion{83453:83671}\def\cmsCernNo{2011-177}\def\cmsCernDate{\today}\def\cmsMessage{Submitted to the Journal of High Energy Physics}

\begin{document}\cmsNoteHeader{BPH-10-014}

\hyphenation{had-ron-i-za-tion}
\hyphenation{cal-or-i-me-ter}
\hyphenation{de-vices}
\RCS$Revision: 83671 $
\RCS$HeadURL: svn+ssh://alverson@svn.cern.ch/reps/tdr2/papers/BPH-10-014/trunk/BPH-10-014.tex $
\RCS$Id: BPH-10-014.tex 83671 2011-11-07 12:23:27Z alverson $

\newcommand{\Jpsi}{\JPsi}
\newcommand{\Psis}{\Pgy}

\newcommand{\ups}{\PgU}
\newcommand{\cc}{\ccbar}
\newcommand{\mypt}{{\ensuremath{p_{\mathrm{T}}}}}
\newcommand{\ptjpsi}{{\ensuremath{p_{\mathrm{T}}^{J/\psi}}}}
\newcommand{\bb}{\bbar}
\newcommand{\etajpsi}{{\ensuremath{y}}}
\newcommand{\ptmu}{{\ensuremath{p_{\mathrm{T}}^{\mu}}}}
\newcommand{\etamu}{{\ensuremath{\eta^{\mu}}}}
\newcommand{\dsigmadpt}{{\ensuremath{\mathrm{d}^2\sigma/(\mathrm{d}p_\mathrm{T} \mathrm{d}y)}}}
\newcommand{\ptb}{{\ensuremath{p_T^{H_b}}}}
\newcommand{\effic}{{\ensuremath{\epsilon}}}

\cmsNoteHeader{BPH-10-014} 
\title{\texorpdfstring{$J\!/\!\psi$ and $\psi$(2S) production in pp collisions at $\sqrt{s} = 7\TeV$}{J/psi and psi(2S) production in pp collisions at sqrt(s) = 7 TeV}}

\date{\today}

\abstract{
A measurement of the $J\!/\!\psi$ and $\psi\rm{(2S)}$
production cross sections
in pp collisions at $\sqrt{s}=7$~TeV  with the CMS experiment at the LHC
is presented.
The data sample corresponds to an integrated luminosity of
37~pb$^{-1}$. Using a fit to the invariant mass and decay length
distributions, production cross sections
have been measured separately for prompt and non-prompt charmonium
states, as a function of the meson transverse
momentum in several rapidity ranges.
In addition, cross sections restricted to the acceptance of the CMS
detector are given, which are not affected by the polarization of the
charmonium states. The ratio of the differential production cross sections of
the two states, where systematic uncertainties
largely cancel, is also determined.
The branching
fraction of the inclusive $\mathrm{B} \to \psi(\mathrm{2S}) X$
decay is extracted from the ratio of the non-prompt cross
sections to be:
$$
\mathcal{B}(\mathrm{B} \to \Psis X) = (3.08 \pm 0.12\,(\text{stat.+syst.}) \pm 0.13\,(\text{theor.}) \pm 0.42 \,(\mathcal{B}_\text{PDG})) \times 10^{-3}.
$$
}

\hypersetup{%
pdfauthor={CMS Collaboration},%
pdftitle={J/psi and psi(2S) production in pp collisions at sqrt(s) = 7 TeV},%
pdfsubject={CMS},%
pdfkeywords={CMS, physics, software, computing}}

\maketitle

\section{Introduction\label{sec:intro}}
Quarkonium production at hadron colliders provides important
tests of calculations in the context of
both perturbative and non-perturbative quantum chromodynamics (QCD),
via measurements of
production cross sections and polarizations.

The \Jpsi~and \Psis~mesons can be produced in proton-proton (\Pp\Pp) collisions through
two mechanisms: prompt mesons directly produced in the
primary interaction
and non-prompt mesons from the decay of directly produced b hadrons.
In addition, \Jpsi~production can also occur via decays of heavier charmonium
states, both S-wave (the \Psis~itself) and P-wave (the three $\chi_c$ states).
The determination of the latter contribution is
challenging both theoretically and experimentally, because it requires
detection of the low-energy photons from $\chi_{c}$ decays.
In nonrelativistic QCD (NRQCD)
models~\cite{Artoisenet:2007xi,Artoisenet:2007qm}, by adding a
contribution to
prompt charmonium production through colour-octet states~\cite{Chao},
a satisfactory description of prompt \Jpsi~and \Psis~meson cross
sections at the Tevatron~\cite{cdfxs_psip} has been obtained. However, in
these calculations the fraction of \Jpsi~originating from $\chi_{c}$ decays
must be assumed from experimental measurements with large uncertainties,
which makes the \Psis~mesons cleaner probes of NRQCD predictions.

Non-prompt \Jpsi~and \Psis~production can be directly related to b-hadron
production,
providing a measurement of the b-hadron cross section in pp collisions.
Past discrepancies between the Tevatron results and the next-to-leading-order
(NLO) QCD calculations have been resolved using the fixed-order
next-to-leading-log (FONLL)
approach and updated measurements of the $\cPqb \rightarrow \Jpsi$
fragmentation and decay~\cite{cacciar1,cacciar2}.

Measurements of the prompt and non-prompt production cross
sections
of \Jpsi\/ mesons decaying to muon pairs
was published using the first Compact Muon Solenoid (CMS) data~\cite{JPsiPaper},
as well as data from other LHC experiments~\cite{atlas,lhcb,alice}.
 The present analysis extends the CMS result with a larger amount of
statistically independent
data and  complements it by providing a measurement of the \Psis~production
cross section, as well as the ratio of the two cross sections.
Higher trigger thresholds induced by the increased LHC luminosity
do not allow the current
measurement to reach charmonium transverse-momentum (\mypt ) values as low as in Ref.~\cite{JPsiPaper}, but
the high-\mypt~reach is increased by the much larger amount of data.
The advantage
of measuring the \Psis~to \Jpsi~cross-section ratio
lies in the cancellation of several experimental and theoretical uncertainties.

The polarizations of the \Jpsi~and \Psis~states
affect the muon momentum spectrum in the laboratory frame,
thus influencing the charmonium acceptance and, as a consequence,
the extracted cross sections.
Therefore
it was decided to present the results in two different ways.
The first approach
assumes unpolarized production for prompt \Jpsi~and \Psis\
whereas non-prompt mesons are assumed to
have the polarization
generated by the \textsc{EvtGen}
Monte Carlo program~\cite{bib-evtgen}, corrected to match the most recent measurements~\cite{babarpol}.
Typical changes of the measured
cross sections resulting from using hypotheses
of full longitudinal or full transverse polarizations are also given.
The second approach provides
results restricted to the phase-space region of the
CMS muon detector
acceptance, in order to avoid corrections which
depend on the unknown polarizations of the
two charmonium states.

The paper is structured as follows. In Section~\ref{sec:cms}, a brief
description of the CMS detector is provided. In Section~\ref{sec:sel}, the data
and Monte Carlo samples are presented, and the event selection is described,
while Section~\ref{sec:inclusive} presents the method to extract the total
\Jpsi~and \Psis~yields.
In Section~\ref{sec:Acc_eff}, corrections for acceptance and efficiency are
explained, which are used to determine the inclusive cross sections,
as discussed in Section~\ref{sec:inclu}. Section~\ref{sec:nonprompt} describes
the method to extract the \Jpsi~and \Psis~non-prompt fractions of the total
yields. In Section~\ref{sec:resu}, the results of the
differential cross sections, the non-prompt fractions, the
cross-section ratios, and the inclusive branching
fraction $\mathcal{B}(\PB \to \Psis X)$
from the ratio of the non-prompt cross sections are given.

\section{The CMS detector}\label{sec:cms}

A detailed description of the detector can be found elsewhere~\cite{JINST}.
The central feature of the CMS apparatus, composed of a central barrel and two
endcaps, is a  6~m diameter superconducting
solenoid producing a 3.8~T magnetic field.
Within the magnetic field volume are the
silicon tracker, the crystal electromagnetic
calorimeter, and the brass/scintillator hadron calorimeter.

The
coordinate system adopted by CMS has the origin at the nominal collision point,
the $y$ axis pointing vertically upward, and the $x$ axis pointing radially
toward the centre of the LHC ring. The $z$ axis points along the
anti-clockwise
beam direction defining a right-handed
coordinate system.
The polar angle $\theta$ is measured from the $z$ axis.
The pseudorapidity of a particle
is defined as $\eta = - \ln [\tan (\theta / 2)]$, which approaches the rapidity
 $y = 0.5 \ln[(E+cp_z)/(E-cp_z)]$ in the ultra-relativistic limit,
where $E$ and $p_z$ are the particle's energy and longitudinal momentum.

Muons are detected in the pseudorapidity  range $|\eta|< 2.4$
by three types of gas-based detectors
embedded in the steel return yoke:
drift tubes in the barrel, cathode
strip chambers in the endcaps, and resistive plate chambers in both the
barrel and endcaps.

The silicon tracker consists of the inner pixel-based detector followed
by layers of microstrip detectors.
The strong magnetic field and the
good position resolution of the silicon tracker enable the
transverse momentum of muons matched to reconstructed
tracks to be measured with a resolution of $\sim$\,1.5\,\%  for
\mypt~smaller than 100\GeVc.

The first level (L1) of the CMS trigger system, composed of custom
hardware processors, uses information from the calorimeters and muon
detectors to select the most interesting events.
The high level trigger (HLT) runs on a processor farm to
reduce further the rate before data storage.

\section{Data selection and event reconstruction}\label{sec:sel}

This analysis is based on
a data sample collected in 2010
with the CMS detector, in \Pp\Pp\ collisions
at a centre-of-mass
energy of 7\TeV.
The sample is
selected to have consistent trigger
requirements for the data used in the analysis and without overlap
with the sample used in Ref.~\cite{JPsiPaper}. It
corresponds to a total integrated luminosity of
$36.7 \pm 1.5\pbinv$~\cite{bib-lumi}.
During this data-taking period,
there were on average
2.2  inelastic
\Pp\Pp\ collisions per bunch crossing at the CMS interaction
region.

The \Jpsi~and \Psis~mesons are reconstructed in the
$\mu^+\mu^-$ decay channel.
This analysis is based on events selected by
dimuon triggers that exploit advanced processing at the HLT level.
Information from all
three muon systems, as well as from the tracker, are used
to make the trigger decision.
Both muons are required to be consistent with
a L1 muon signal, requiring at least
two independent segments in the muon chambers,
and to be matched to a
track reconstructed in a region of interest defined by the L1 seed.
No explicit
requirement on the transverse momentum \mypt~is applied.

Simulated events are used to tune the selection criteria,
check the agreement with data, compute the acceptance, and
derive efficiency
corrections, as well as for systematic studies.
Prompt \Jpsi~and \Psis\ events
are simulated using \PYTHIA\ 6.422~\cite{bib-PYTHIA},
which generates events based on the leading-order colour-singlet and colour-octet mechanisms.  Colour-octet states undergo a shower evolution. We use the NRQCD matrix element tuning obtained by fitting NRQCD calculations to CDF data~\cite{kramer,marianne}.
In the absence of consistent theoretical and experimental information
about the \Jpsi~and \Psis\ polarizations,
the dilepton decay distribution is assumed to be isotropic.
Simulated events with b-hadron decays are also generated with \PYTHIA , and the b hadrons
are forced to decay
inclusively into \Jpsi~and \Psis~using the \textsc{EvtGen} package.
Photon final-state
radiation (FSR)
is  implemented using \PHOTOS~\cite{bib-photos1,bib-photos2}.

The off-line event selection, only briefly summarized here, is very
similar to the one used in Ref.~\cite{JPsiPaper}.
Muon candidates are reconstructed from the combination of
muon-detector and silicon-tracker hits.
The muons are required to pass the following
criteria in the tracker:
have at least 10 tracker hits,
at least two of which are required to be in the pixel layers;
have a $\chi^2$ per degree of freedom smaller than
1.8; and
pass within a cylinder of radius 3~cm and length 30~cm
centred at the beam-spot centroid position
and parallel to the beam line.

To select
events with \Jpsi~or \Psis~decays, muons with opposite charge are paired and their
invariant mass is computed. The mass
is required to be between
2.5 and 4.7\GeVcc.
The two muon trajectories are refitted with a common vertex constraint, and events are retained
if the $\chi^2$ probability of the fit is larger than 1\%. If more than one muon
pair is found in the event,
the one with the largest
vertex $\chi^2$ probability
is retained.

The dimuon L1 triggers include a veto,
whose specific criteria depend on the type of muon chamber and the region
 of the detector: this rejects muon signals whose spatial separation in the muon stations is too small, in order to avoid spurious dimuon signatures
from
a single muon. As a consequence, the dimuon
sample is split in two,
depending on the signed difference in azimuthal angle ($\Delta\phi$)
between the positively and the negatively charged muons.
Muons that bend towards each other in the magnetic field are called type-C (``convergent'') dimuons, while  muons
that bend away from each other are type-D (``divergent'') dimuons.

Dimuons of type D are
much less affected
by the trigger veto, while dimuons
of type C may cross
at the muon stations. This
causes
sizeable correlations between the two muon detection efficiencies,
and this effect is larger in the forward region.
Therefore, in addition to the above requirements,
all type-C dimuons are rejected
for the inclusive cross-section measurements.
This corresponds to a 48\% reduction in the yield.
For the non-prompt fraction
determination (which is largely efficiency independent), type-C dimuons are
only rejected in the high dimuon rapidity region $1.6 < |y| < 2.4$.

The momentum measurement of charged tracks in the CMS detector
has systematic uncertainties that are due to imperfect knowledge
of the magnetic field, modelling of the detector material,
and sub-detector misalignment. These effects can
shift and
change the width of the
mass peaks of dimuon resonances.
In addition to calibrations already applied to the data \cite{bib-magneticfield,bib-material,bib-trackeralignment},
residual effects are determined by studying the dependence of the reconstructed dimuon peak shapes
on the muon kinematics,
as was done in Ref.~\cite{JPsiPaper}.

Because of
the large difference in the branching fractions to dimuons
of the two states,
the measured \Psis~yield is much smaller than the \Jpsi\ yield.
For this reason, different binnings are used for the
differential cross sections in
\mypt\ and $|y|$.

\section{Inclusive yield determination \label{sec:inclusive}}

Two methods are used
to extract the inclusive yields
from  the $\mu^+\mu^-$ invariant mass distribution,
either fitting the \Jpsi~peak alone in a restricted mass window,
or fitting the combined \Jpsi~and \Psis~distribution. Yields are derived
using an extended unbinned max\-imum-\-like\-li\-hood meth\-od.

\begin{figure}[htbp!]
\centering
{
\includegraphics[width=6.5cm,angle=90]{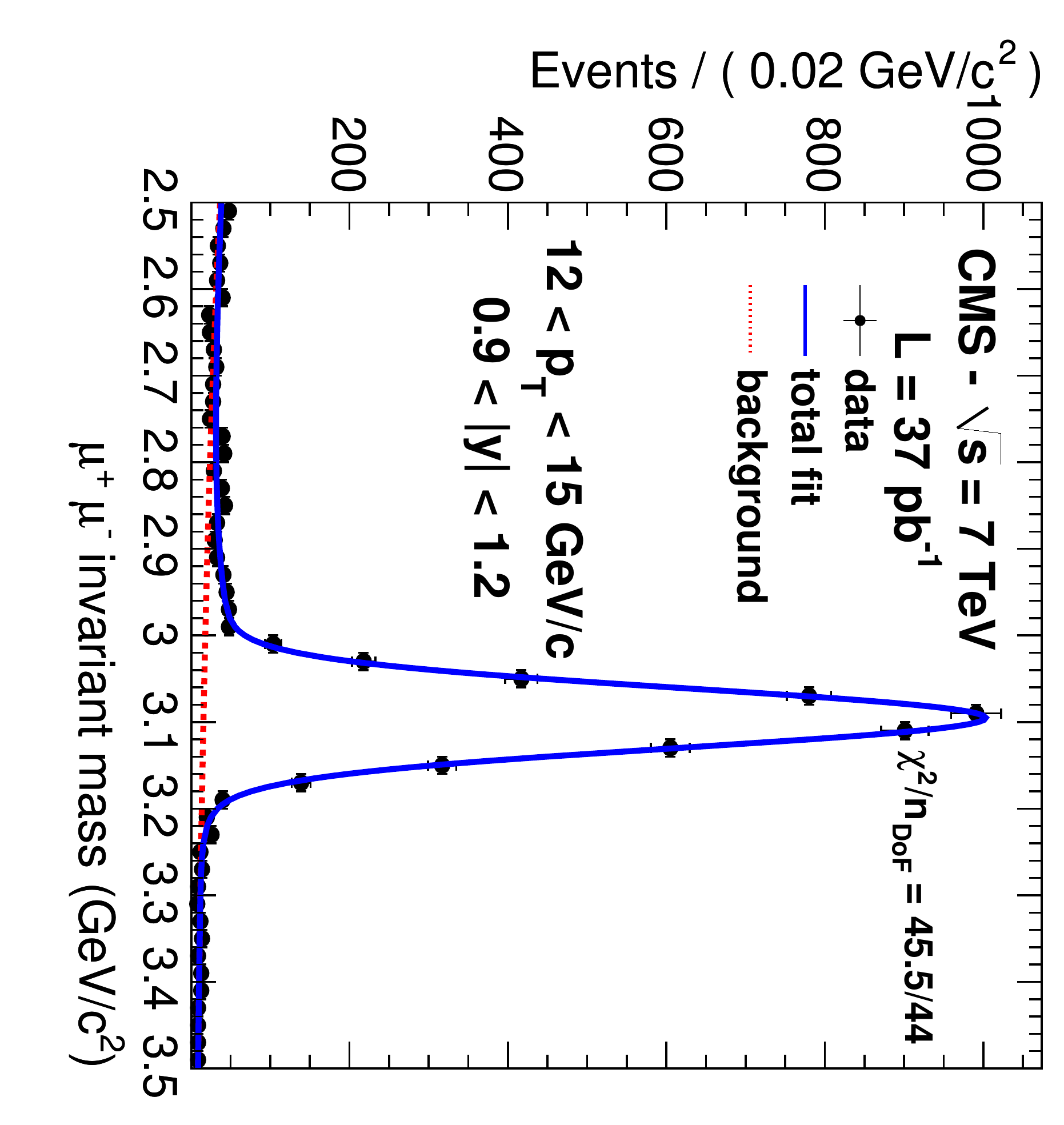}
\includegraphics[width=6.8cm,angle=90]{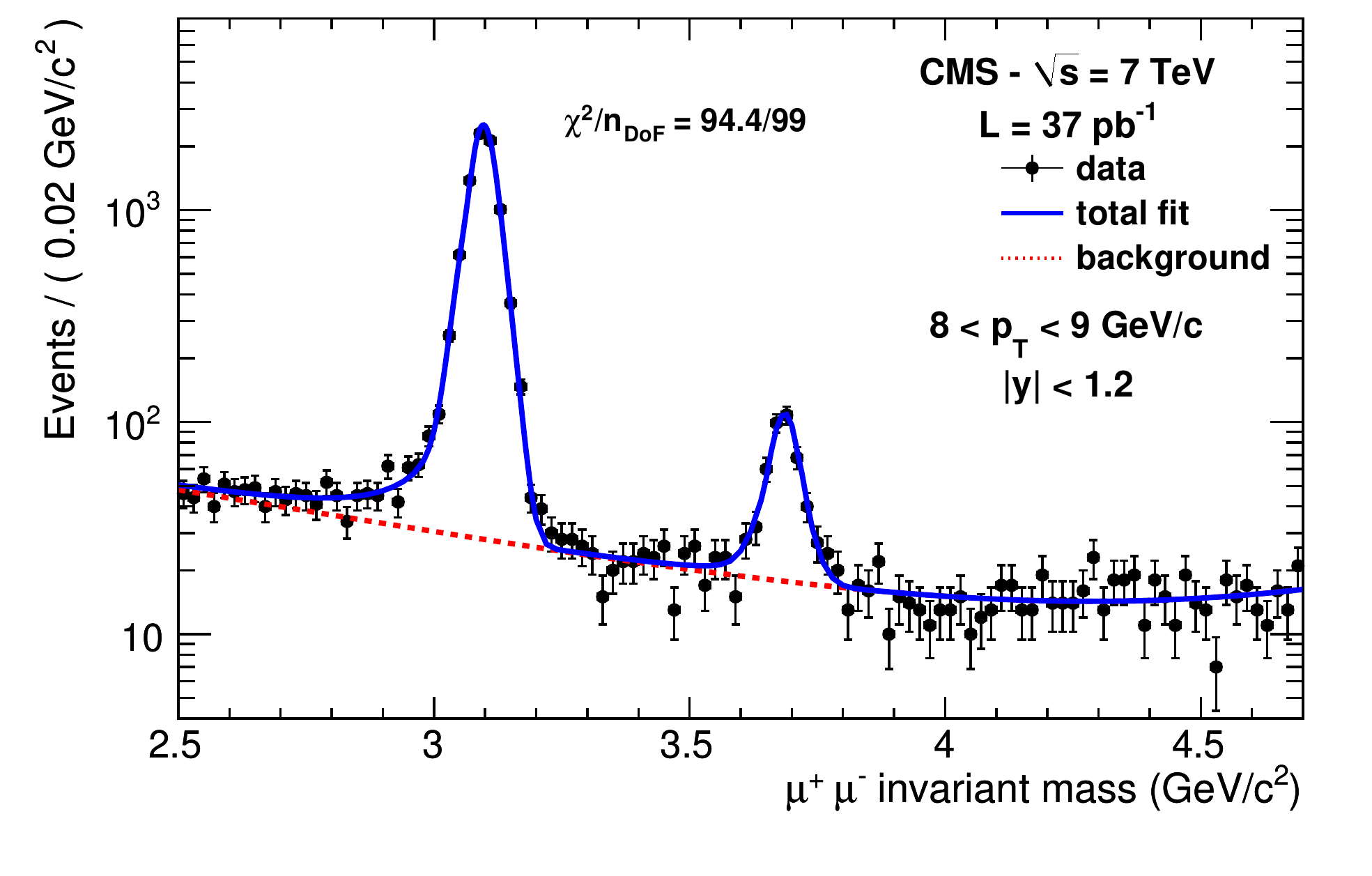}
}
\caption{
Top: The $\mu^+\mu^-$ invariant mass distribution in the \Jpsi\ region and the result of the
fit for the bin: $0.9 < |y| < 1.2$, $12 < \mypt < 15$
\GeVc. Bottom:  \Jpsi~and \Psis~mass distribution fit for the bin:  $|y| < 1.2$, $8 < \mypt < 9\GeVc$.
The solid and dashed lines represent the total fits and their background components, respectively.
}
\label{fig:allmassfits}
\end{figure}

In both types of fits, the sum of a Gaussian and a
Crystal Ball~\cite{bib-crystalball} function is used for
the description of
the signal, simultaneously taking into account FSR
and rapidity-dependent resolution variations.
In the \Jpsi-only fits, an exponential function
is used to describe 
the background.
Figure~\ref{fig:allmassfits} (top) shows an example of a fitted mass distribution. The mass resolution ranges from about $20\MeVcc$ in the low rapidity region to
$35\MeVcc$ at intermediate rapidities, up to about $50\MeVcc$ in
 the forward, high rapidity region.

In
the second type of
fits,
the two mass peaks
and the background 
are fitted
simultaneously. 
For the \Jpsi~and \Psis~signal peaks,
the same probability density function (pdf) is used,
with
the following
constraints on the parameters:
the ratio of the central values of the two masses is fixed to the
world average value~\cite{bib-pdg};   
the
widths, scaled by the nominal mass values, are constrained to be the same,
and
the
parameters
describing the asymmetric tail of the Crystal Ball function
are constrained to be equal.
The background is modelled by two exponentials.
Figure~\ref{fig:allmassfits} (bottom) shows
an example of
a fitted mass distribution.

The systematic uncertainties on the mass distribution fits are
estimated
by changing the analytical form of the
signal and background pdf hypotheses (a single Crystal Ball function is used
for the signal and polynomial pdfs for the background) in the two types of
fits.
The largest variation
in
the yield
of each fit
is taken as
the 
systematic uncertainty.

\section{Acceptance and efficiency}\label{sec:Acc_eff}
As  discussed in Section~\ref{sec:intro}, measurements will be
presented
using two different approaches. In the first approach,
the observed number of \Jpsi~events is corrected for the detector acceptance and
reconstruction efficiency in every bin in which the cross section is measured. As the acceptance is strongly dependent on the assumed polarization of the charmonium state,
in the second approach we
provide measurements exclusively within the CMS
detector acceptance, where only detector
efficiency corrections are made
and without any polarization-related uncertainties.

\subsection{Acceptance}

The acceptance reflects the geometrical coverage of the CMS detector
and the kinematic reach of the muon trigger and reconstruction,
constrained by the amount of
material in front of the muon detectors and by the track curvature
in the magnetic field.

In the simulation, both muons are required to
be within the geometric acceptance of the muon
detectors.
A single muon is defined as detectable if it satisfies the following requirements at generator level:
\begin{eqnarray}
\nonumber
\ptmu>4.0~\GeVc & \textrm{ for } & |\etamu|<1.2 \\
\ptmu>3.3~\GeVc & \textrm{ for } & 1.2<|\etamu|<2.4 \quad.
\label{eq:muacc}
\end{eqnarray}

The \Jpsi~acceptance $A$ is defined as the fraction of detectable
$\Jpsi\to\mu^+\mu^-$ decays,
as a function of the generated dimuon transverse momentum \mypt~and rapidity $y$,

\begin{equation}
A(\mypt, y;\lambda_\theta)=\frac
{N_{\mbox{\small{det}}}(\mypt, y;\lambda_\theta)}
{N_{\mbox{\small{gen}}}(\mypt, y;\lambda_\theta)} \quad,
\label{eq:acceptance}
\end{equation}
where
$N_{\mbox{\small{det}}}$
is the number of detectable \Jpsi~events in a given (\mypt, $y$) bin,
and  $N_{\mbox{\small{gen}}}$
is the corresponding total number of generated \Jpsi~events in the Monte Carlo (MC) simulation. An analogous definition holds for \Psis.
The parameter $\lambda_\theta$ reflects the fact that the acceptance is
computed for various polarization scenarios,
which lead to different muon spectra in the laboratory frame.

For the acceptance calculation, a dedicated sample of generated events is used,
with no restrictions on the phase space.
The large number of simulated events
allows a much smaller bin size
for determining $A$ with respect
to that  used for the cross-section
determination in data.

To study the effect of \Jpsi\ and \Psis\ polarization on the acceptance,
these events are reweighted
depending on the values of the polar and azimuthal
angles as computed in two different frames (helicity and Collins-Soper~\cite{bib-faccioli}).
The angular distribution for the decay of a $J=1$ state into fermions
is used, which is a function of three independent parameters
$\lambda_\theta$, $\lambda_\phi$, and $\lambda_{\theta\phi}$:
 \begin{equation}
 W(\cos\theta,\phi) =
\frac{3}{2(3+\lambda_\theta)}\cdot(1+\lambda_\theta \cos^2\theta + \lambda_\phi
 \sin^2\theta \cos2\phi + \lambda_{\theta\phi}\sin2\theta\cos\phi) \quad.
 \label{eq:polarization}
 \end{equation}
The choice of zero for all $\lambda$ parameters
corresponds to an unpolarized decay, while $\lambda_\theta=-1$ and $\lambda_\theta=+1$ correspond to fully longitudinal and fully transverse polarizations, respectively.

By default, the prompt \Jpsi~and \Psis~are
assumed to be unpolarized, while the non-prompt mesons are assumed to be
polarized as generated by
{\sc EvtGen}
and corrected to match recent measurements,
as mentioned in the Introduction.
Typical changes of the measured
cross sections when using
alternative polarization scenarios are
 provided in Section~\ref{sec:resu}.

Several sources of systematic uncertainty
on the acceptance
have been investigated:

\begin{itemize}
\item \emph{Kinematic spectra}.
Different
\mypt\ and $y$
spectra of the generated \Jpsi\ and \Psis\ might produce different acceptances,
as the acceptance is defined by single-muon criteria.
Spectra from theoretical predictions presented in Section~\ref{sec:resu} have been used to recompute the acceptance,
and the difference from that obtained with the \PYTHIA\ spectrum
has been taken as a systematic uncertainty.

\item \emph{Final-state radiation}.
The generated dimuon momentum may differ from the \Jpsi~and \Psis~momentum,
because of FSR. The difference between the acceptance computed using the dimuon or the
charmonium
variables in Eq.~(\ref{eq:acceptance}) is taken as a systematic uncertainty.

\item \emph{\PB\ polarization}.
The \Jpsi\ and \Psis\ mesons  produced in \cPqb-hadron decays have a
different acceptance
with respect to the prompt ones. The corresponding systematic uncertainty
is evaluated by taking the
difference of the default choice
(corrected to match the experimental results of Ref.~\cite{babarpol})
with respect to the one predicted by {\sc EvtGen}.

\item \emph{\mypt\ calibration}.
The muon transverse momenta in data have been calibrated as
described in Ref.~\cite{JPsiPaper}.
A difference in the momentum resolution between data and simulated events would also give a systematic uncertainty on the acceptance.
The acceptance has been computed with simulated muon momenta smeared according
 to the resolution
measured
with data~\cite{bib-trackermomentum}.
The uncertainty on the measured resolution
was
used to apply an additional smearing on the simulated momenta;
the acceptance has been recalculated and the shift
taken as a systematic uncertainty.
\end{itemize}

\subsection{Muon efficiency}

The single-muon efficiency
 is measured from data
for muons in the acceptance,
as described in Refs.~\cite{bib-muonreco,bib-trackingefficiency},
and is based on the
\textit{tag-and-probe} (T\&P) method. For this purpose,
independent sets of triggers
are used for which online
requirements either on the muon or the tracker tracks are not applied,
thus yielding samples which are unbiased with respect to the corresponding
selections.

The combined trigger and offline reconstruction efficiency for a single muon is defined as:

\begin{equation}
\effic(\mu)  =
\effic_{\mbox{\small trig $|$ off}}
\cdot \effic_{\mbox{\small off $|$ ID}}
\cdot \effic_{\mbox{\small ID $|$ track}}
\cdot \effic_{\mbox{\small track}},
\end{equation}

where $ \effic_{\mbox{\small track}} $ is the offline
tracking efficiency,
$ \effic_{\mbox{\small ID $|$ track}} $ refers to the muon identification in the muon systems for a tracker-reconstructed muon,
$ \effic_{\mbox{\small off $|$ ID}} $ refers to the specific quality requirements applied to reconstructed muons,
and $ \effic_{\mbox{\small trig $|$ off}}$ is the probability for an offline reconstructed muon to have also fired the trigger.

The muon identification and trigger efficiencies
($\effic_{\mbox{\small trig $|$ off}}$,
$\effic_{\mbox{\small off $|$ ID}}$ and
 $\effic_{\mbox{\small ID $|$ track}}$)
have the strongest
\ptmu~and $|\etamu|$ dependence and are
determined
in 15 bins of \ptmu\ ($3.3 < \ptmu < 50\GeVc$) and 14 bins of $|\etamu|$  ($0 < |\etamu| < 2.4$),
allowing an adequate description of the turn-on efficiency curves.
Since the tracking efficiency is almost constant for this momentum and rapidity range, broader bins are used.

The efficiency to detect a dimuon event is expressed as:
\begin{equation}
\epsilon (\mu^+\mu^-) = \epsilon ({\mu^+})  \cdot \epsilon ({\mu^-}) \cdot \rho
\cdot \epsilon_{\mbox{\small vertex}},
\label{eq:jpsi_eff}
\end{equation}
where
$\epsilon ({\mu^+}) $ and $\epsilon ({\mu^-}) $ are the single-muon efficiencies, and
$\epsilon_{\mbox{\small vertex}}$ is the efficiency of the vertex
$\chi^2$ requirement,
calculated from the data by determining the yields in
regions of
large \mypt\ and $|y|$  by alternatively applying and not applying this
requirement.
The $\rho$ factor,
defined by Eq.~(\ref{eq:jpsi_eff}), represents a correction
to the efficiency factorization hypothesis: it accounts
for the finite size of the $(\ptmu, \etamu)$ bins and, more importantly, for
the possible bias introduced by the T\&P measurement, due to correlation effects
as discussed in Section~\ref{sec:sel}.
In order to determine $\rho$, the efficiencies
have also been evaluated using T\&P techniques
on simulated events and their product has been compared with the true dimuon
efficiency.
Except for some bins at high \mypt, the values are
found to satisfy $|1-\rho| < 10\%$.

For the acceptance-corrected cross-section results, the
acceptance and efficiencies are combined into a single
factor, which is computed
for each (\mypt, $y$) bin and is defined as:

\begin{equation}
\bigg\langle \frac{1}{A \cdot \epsilon} \bigg\rangle_\text{bin}
\equiv
\frac{1}{N_{\rm event}}\sum_{k=1}^{N_{\rm event}} \frac{1}{A_k \cdot \epsilon_k(\mu^+\mu^-)} ,
\end{equation}
where the average is taken over the
data events in each
bin, using the ``fine-grained'' bins of the acceptance and
the event-by-event efficiency obtained from the single-muon efficiencies using
Eq.~(\ref{eq:jpsi_eff}).

Similarly, for the results which are not corrected for acceptance, the
efficiency factor is determined as:

\begin{equation}
\bigg\langle \frac{1}{\epsilon} \bigg\rangle_\text{bin}
\equiv
\frac{1}{N_{\rm event}}\sum_{k=1}^{N_{\rm event}} \frac{1}{\epsilon_k(\mu^+\mu^-)} .
\end{equation}

Two sources of systematic uncertainty in the efficiency are considered:
\begin{itemize}
\item The uncertainties on the measured muon efficiencies propagate as
systematic errors on the cross section
measurement through the correction factor
$\langle \frac{1}{A \cdot \epsilon} \rangle_\text{bin}$ (or $\langle \frac{1}{\epsilon} \rangle_\text{bin}$).
The effect has been estimated on a statistical basis in each bin by performing Monte Carlo
pseudo-experiments,
in which the muon efficiencies were varied randomly according to
a probability density
built by joining the left and the right side of two Gaussians with different
widths, in order to allow
for asymmetric errors. The r.m.s. of these correction factors in each bin has been taken
as the systematic uncertainty
associated with the
single-muon efficiency.
\item The full difference $|1-\rho|$ in each bin is taken as a
systematic uncertainty due to the efficiency correlation.
\end{itemize}

\section{Inclusive cross-section determination}\label{sec:inclu}

The
inclusive double-differential cross section is
given by:
\begin{equation}
\frac{\mathrm{d}^2\sigma}{\mathrm{d}\mypt \mathrm{d}y}(\Jpsi) \cdot \mathcal{B}(\Jpsi\rightarrow \mu^{+}\mu^{-})=\frac{N^{\text{corr}}_{\Jpsi}(\mypt , |y|)}{\int{}
L \, \mathrm{d}t\cdot\Delta \mypt \cdot\Delta y}\quad ,
\end{equation}
where
$\int{}L \, \mathrm{d}t$ is the integrated luminosity, $\Delta \mypt$ and $\Delta y$ are
the \mypt~and $y$ bin widths, $\mathcal{B}(\Jpsi\rightarrow \mu^{+}\mu^{-})$ is
the decay branching
fraction
of the \Jpsi~into two muons, and
$N^{\text{corr}}_{\Jpsi}(\mypt, |y|)$ is the corrected \Jpsi~yield in a given
(\mypt, $|y|$) bin.
The corrected yield is obtained from the fitted signal
yield $N_{\Jpsi}$ via
$N^{\text{corr}}_{\Jpsi} =  N_{\Jpsi} \cdot \langle \frac{1}{A \cdot \epsilon} \rangle_\text{bin}$ in the case where the \Jpsi~yields are corrected
for acceptance and
efficiency, and  $N^{\text{corr}}_{\Jpsi} =  N_{\Jpsi} \cdot \langle \frac{1}{\epsilon} \rangle_\text{bin}$  in the case where the results are uncorrected
for acceptance.
An analogous formula applies for the $\Psis$ double-differential cross section.

\begin{figure}[htpb!]
\centering
{
\includegraphics[width=10cm]{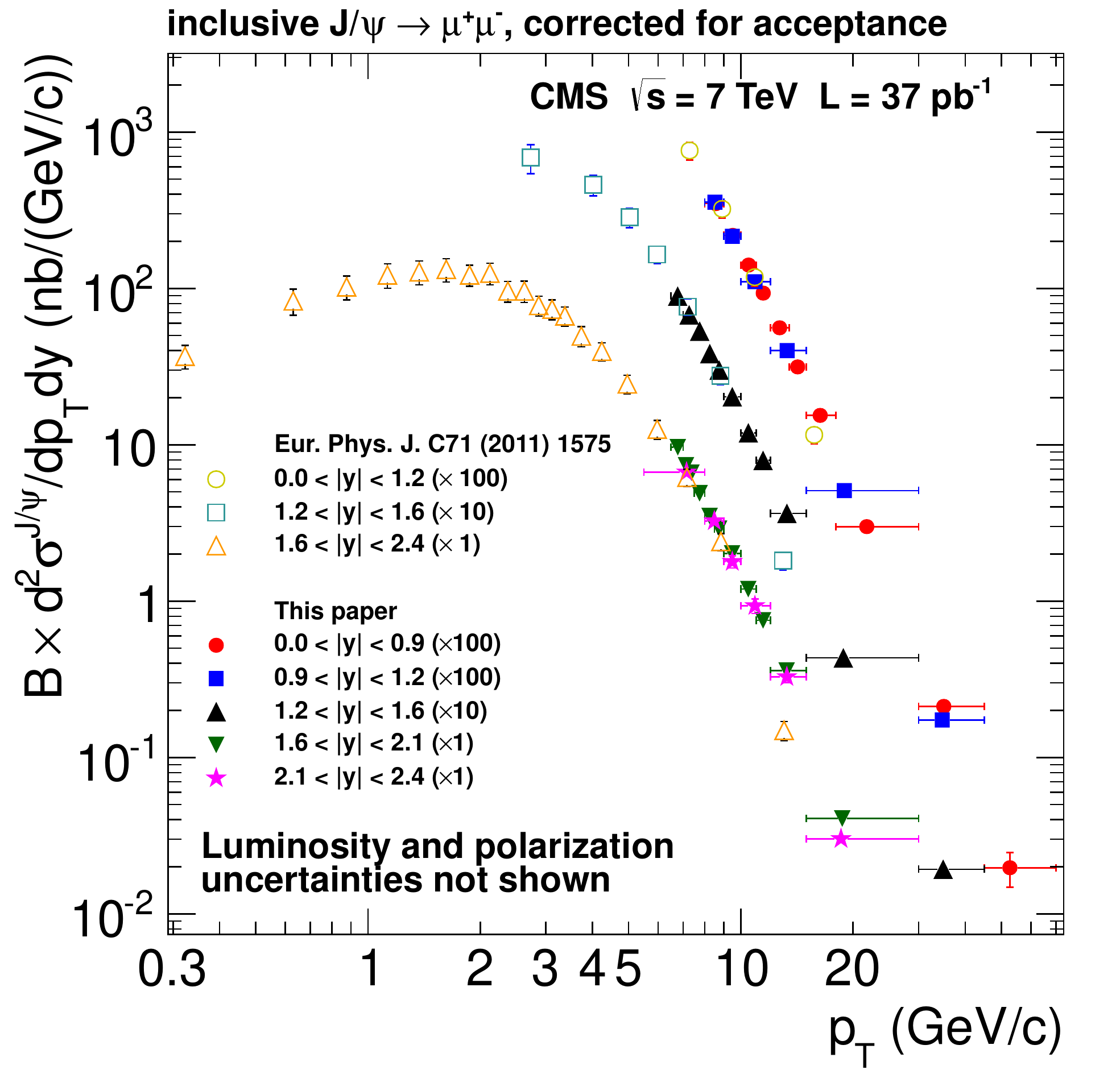}
}
\caption{Measured differential cross section for \Jpsi~inclusive production as a function of \mypt~for five rapidity bins, fully corrected for acceptance
and efficiency.
Also plotted are the results published in Ref.~\cite{JPsiPaper}, which extend to a lower \mypt~range. The error bars on data points include all the statistical and systematic contributions except luminosity and polarization.
The measurements have been offset by the numerical values given in the legend for easier viewing.
}
\label{fig:XsecJpsiOld}
\end{figure}

Figure~\ref{fig:XsecJpsiOld}
shows  the measured (fully corrected) \Jpsi~inclusive cross section as a function of \mypt~for the various rapidity bins. They are compared with our previous results published in Ref.~\cite{JPsiPaper},
which are statistically independent and remain of interest since they partially
overlap with the present results and cover a lower \mypt~range.
A good agreement is observed. In this figure, as well as in the cross-section plots of Section~\ref{sec:resu}, multiplicative factors -- appearing as additive offsets on the
log scale -- are used to
achieve a convenient graphical separation of the measurements from different rapidity bins.

\section{Prompt and non-prompt fractions \label{sec:nonprompt}}

To estimate the \Jpsi~fraction from \cPqb-hadron decays, a
two-dimensional fit
is performed,
in which the pdfs and fit procedure are the
same as those described in Ref.~\cite{JPsiPaper}.
The variables used for the two-dimensional fits are the dimuon invariant mass and the
``pseudo proper decay length'' $\ell_{\Jpsi}$, defined
as the most probable value of the
transverse
distance between the dimuon vertex and the primary vertex, corrected
by the transverse
Lorentz boost of the \Jpsi.
As in Ref.~\cite{JPsiPaper}, the primary vertex is chosen as the one
closest to the dimuon vertex in the $z$ direction.

The resolution of the
pseudo proper decay length  is described by a function depending
on an event-by-event uncertainty determined
from the covariance matrices of the
primary and secondary vertex fits. The uncertainty is used as the
r.m.s. of the resolution Gaussian function
that describes
the core of the resolution, while a second Gaussian function with a small
relative normalization (usually $< 1\%$)
parametrizes the effect of incorrect primary vertex assignments.

The pdf $F(\ell_{\Jpsi} ,m_{\mu\mu}, \sigma_\ell)$
for the \Jpsi\ is then:
\begin{eqnarray}\label{eqx3}
\nonumber
F(\ell_{\Jpsi},m_{\mu\mu},\sigma_\ell) & = &
f_{\text{Sig}}\cdot D_{\text{Sig}}(\sigma_\ell)\cdot F_{\text{Sig}}(\ell_{\Jpsi},\sigma_\ell)\cdot
M_{\text{Sig}}(m_{\mu\mu}) + \\
&& (1-f_{\text{Sig}})\cdot D_{\text{Bkg}}(\sigma_\ell) \cdot F_{\text{Bkg}}(\ell_{\Jpsi},\sigma_\ell)\cdot
M_{\text{Bkg}}(m_{\mu\mu}),
\end{eqnarray}
where:
\begin{equation}\label{eqx4}
F_k(\ell_{\Jpsi},\sigma_\ell) = \sum^{2}_{i=1} F_k^{\text{true}}(\ell'_{\Jpsi})\otimes R_i(\ell_{\Jpsi} - \ell'_{\Jpsi}| \mu, s_i \sigma_\ell) .
\end{equation}
and $k =$ \{Sig, Bkg\}. In the equations above:
\begin{itemize}
\item $M_{\text{Sig}}(m_{\mu\mu})$ and $M_{\text{Bkg}}(m_{\mu\mu})$ are the mass pdfs determined for the signal and background in
Section~\ref{sec:inclusive}, and $f_{\text{Sig}}$ is the fraction of signal events in
the entire range of the fit;
\item $F_{\text{Sig}}^{\text{true}}(\ell_{\Jpsi})$ and $F_{\text{Bkg}}^{\text{true}}(\ell_{\Jpsi})$ are the functional forms describing the $\ell_{\Jpsi}$ distribution for the signal and background, respectively. The signal part is given by the sum of prompt and non-prompt components:
$F_{\text{Sig}}^{\text{true}}(\ell_{\Jpsi} )=f_{\cPqb} \cdot f_{\cPqb}(\ell_{\Jpsi}) + (1-f_{\cPqb}) \cdot F_{p}(\ell_{\Jpsi})$,
where $f_{\cPqb}$ is the fraction of $J/\psi$ from b-hadron decays, and
 $F_{p}(\ell_{\Jpsi})$ and $f_{\cPqb}(\ell_{\Jpsi})$ are the $\ell_{\Jpsi}$ distributions for prompt and non-prompt \Jpsi , respectively.
The $\ell_{\Jpsi}$ pdfs for prompt signal and background are the same as in Ref.~\cite{JPsiPaper}. The non-prompt lifetime function is described by an
exponential decay of the b hadron, with a Gaussian smearing function that accounts for the difference between the measured
pseudo proper decay length and the proper decay length of the b hadron;
\item $\sigma_\ell$ is the per-event uncertainty of the decay length and
$D_{\text{Sig}}(\sigma_\ell)$ and $D_{\text{Bkg}}(\sigma_\ell)$ are its distributions
separately for signal and background~\cite{punzibias}. They are obtained
from the signal region of the invariant mass distribution,
after a sideband subtraction, and the
sideband regions, respectively;
\item $R_1$ and $R_2$ represent the core and tail decay-length resolution Gaussian functions: $\mu$ is
their common mean and $s_i$ represent scale factors for the per-event uncertainty,
which are both left free in the fit to account for initial assumptions on
the uncertainties of track parameters. These functions are convolved with $F_k^{\text{true}}(\ell_{\Jpsi})$ to obtain the observed $F_k(\ell_{\Jpsi})$ distributions,
including the experimental resolution ($k =$ \{Sig, Bkg\}).
\end{itemize}

The background is fitted using the events in mass sidebands and the result is used to fix lifetime parameters
of the overall fit in the entire mass region.
The mass sideband region is defined as [2.50, 2.85] and [3.25, 3.35] \GeVcc.

For the determination of the \Psis~non-prompt fraction, the
quantity $\ell_{\Psis}$,
defined as for the \Jpsi~case,
is computed.
In order to constrain the fit and avoid problems due to
limited statistical accuracy,
the \Jpsi~and \Psis~samples are fitted simultaneously
using the same binning as
for the \Psis\ cross-section determination.
The lifetime resolution functions $R_1$ and $R_2$
are constrained
to be described by the same parameters (mean value and scale factors)
and the backgrounds to have the same fractions of long-lived components.

The invariant mass sideband regions used for the determination of the
background parameters
are defined as above for the \Jpsi, and
as [3.35, 3.45] and [3.85, 4.20] \GeVcc for the \Psis.

\begin{figure}[htbp!]
\centering
{
\includegraphics[width=7.5cm]{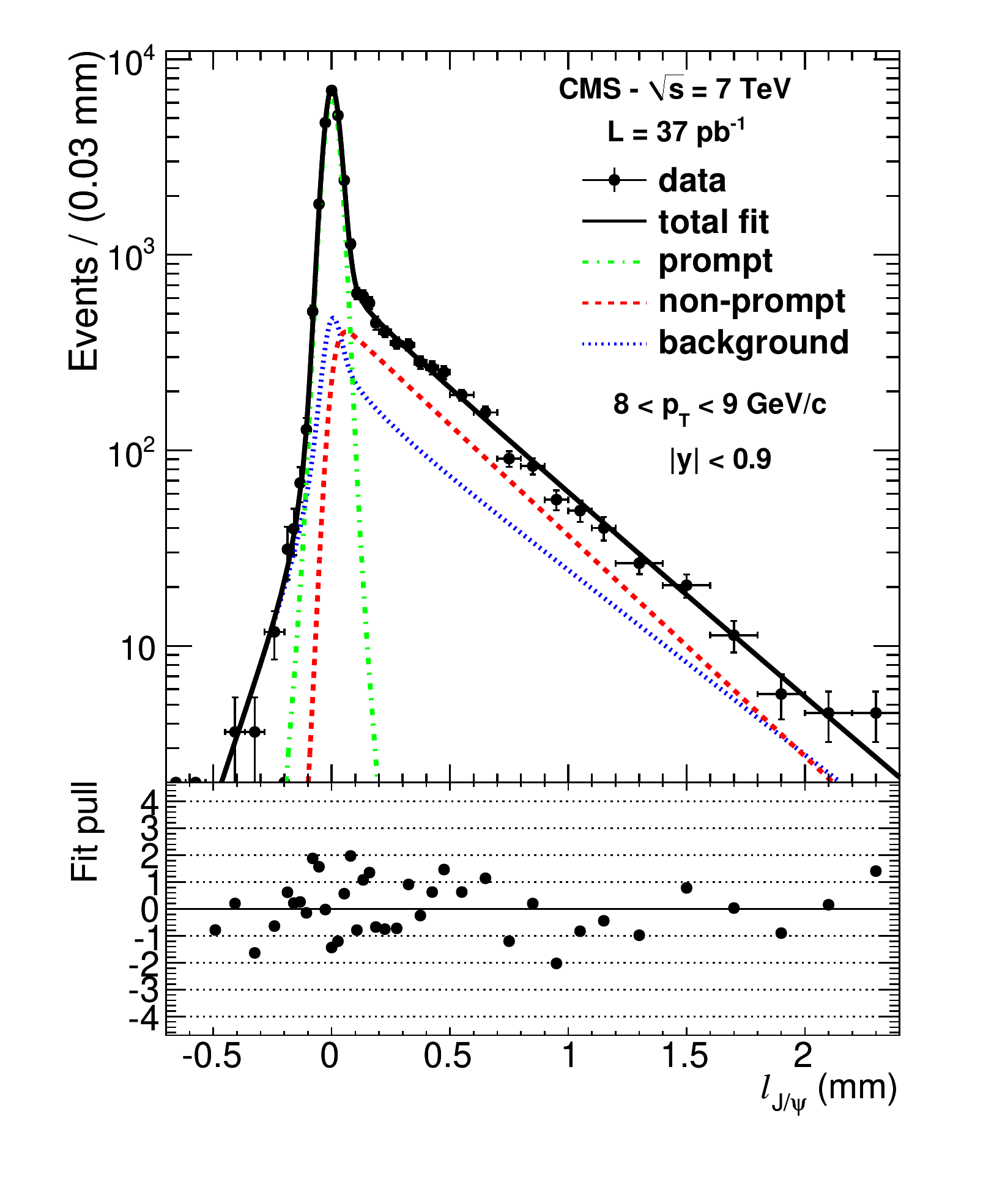}
\includegraphics[width=7.5cm]{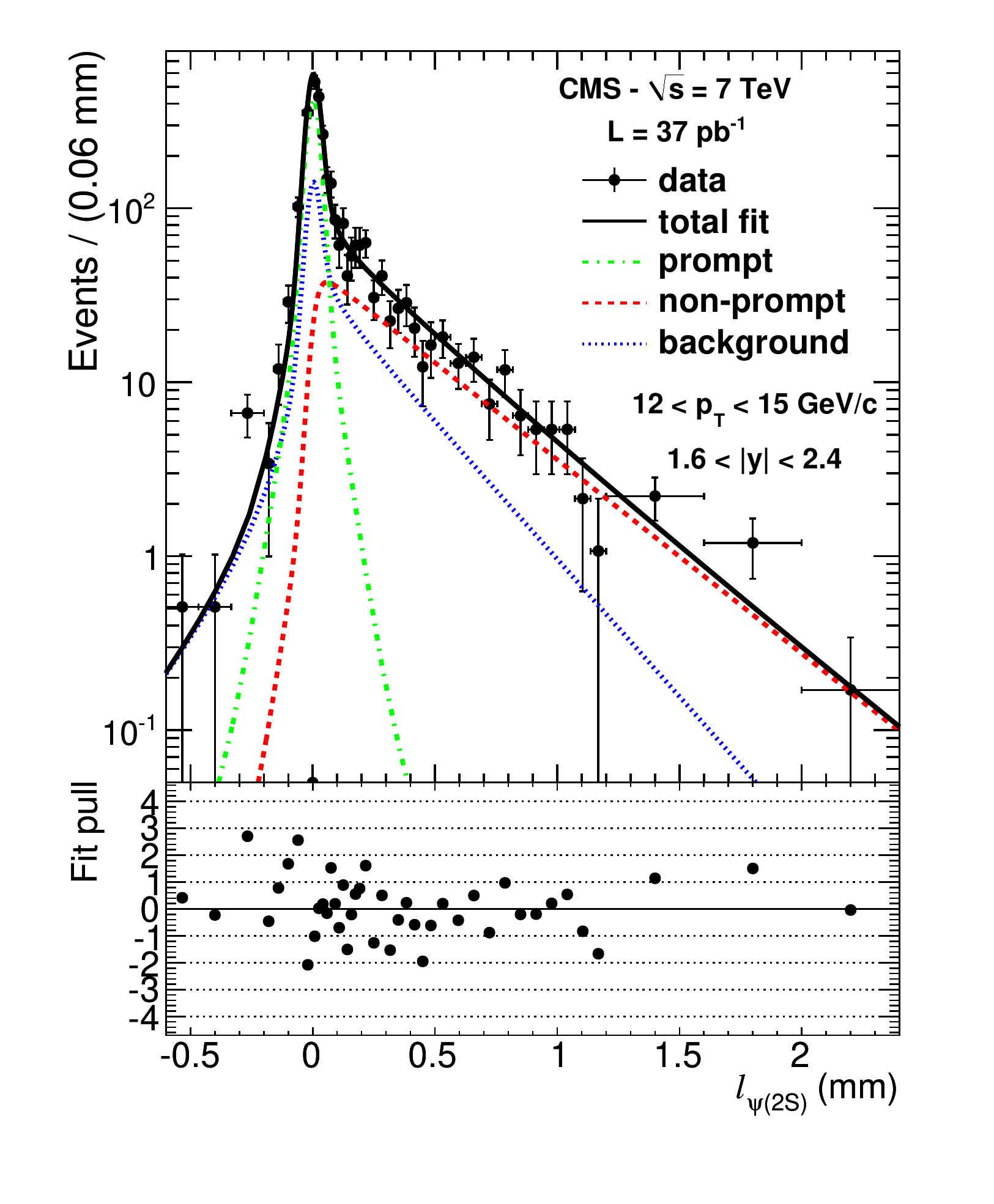}
}
\caption{Left: Projection of a \Jpsi~two-dimensional fit on the $\ell_{\Jpsi}$
dimension in the bin $|y| < 0.9$,
$8 < \mypt < 9\GeVc$ and in the whole mass region [2.50, 3.35] \GeVcc. Right: Projection of a \Jpsi-\Psis~two-dimensional fit on the $\ell_{\Psis}$
dimension in the bin:
$1.6 < |y| < 2.4$, $12 < \mypt < 15\GeVc$, in the \Psis~mass region [3.35, 4.20] \GeVcc. The solid lines represent the total fits; the
prompt, non-prompt and background components are also shown using
green dash-dotted, red dashed and blue dotted lines, respectively. The fit pull plots
show no systematic structures.}
\label{fig:alllifetimefits}
\end{figure}

Figure~\ref{fig:alllifetimefits} shows two
examples of the $\ell_{\Jpsi}$ and $\ell_{\Psis}$ distribu\-tions
with pro\-jec\-tions of two-di\-men\-sion\-al fits on these dimensions,
as well as the prompt
and non-prompt components obtained as described above.
The lower plots in Figure~\ref{fig:alllifetimefits} give the pull
distributions from the fits, and show no systematic structure.

Several sources of systematic uncertainty have been addressed, using mostly the same
procedures as in Ref.~\cite{JPsiPaper}. The main additional systematic effect comes
from attempting to choose
the correct primary vertex of the interaction in the
presence of pile-up.
The sources of systematic uncertainty include the following:
\begin{itemize}

\item \emph{Primary vertex assignment}. In order to estimate the possible
effect of pile-up on the primary vertex estimation,
the primary vertex associated to the dimuon
is chosen as the one with
the largest track $\sum \mypt^2$, instead of the one
closest in $z$ to the dimuon vertex.
The difference between the fitted non-prompt fractions in these two approaches is taken as the systematic uncertainty.

\item \emph{Residual misalignment in the tracker}.
The effect of uncertainties in the measured misalignment of the
tracker modules is estimated by reconstructing the data using different
sets of alignment constants.
The largest difference
in the fit results
with respect to the nominal case
is taken as the systematic uncertainty.

\item \emph{\cPqb-hadron lifetime model}.
An alternative fit method is used, namely the b-hadron lifetime model used in Ref.~\cite{JPsiPaper}, which is based on MC templates; the difference in the fitted
non-prompt
fraction
is taken as the systematic uncertainty.

\item \emph{Background fits}.
The effect of a $\pm 100\MeVcc$ variation in the lower limit of the
low-mass side (upper limit of the high-mass side) of the \Jpsi~(\Psis)
sideband boundaries is taken
as the systematic uncertainty.

\item \emph{Resolution model}.  The nominal
  (double-Gaussian)
  model for the pseudo proper decay length per-event resolution is
  compared with a model using a single-Gaussian shape.  The difference is
  taken as the systematic uncertainty.

\item \emph{Different prompt and non-prompt efficiencies}. The MC simulation predicts slight differences between the prompt and
non-prompt \Jpsi~and \Psis~efficiencies, mostly because of the different
track densities from fragmentation products around the muons.
These are taken into account; the relative difference is propagated to the
non-prompt fraction, and
taken as the
systematic uncertainty.
\end{itemize}

Non-prompt fraction results are given in Section~\ref{sec:resu} and
a
summary of all the systematic uncertainties is given in
Tables~\ref{tab:systsmall} and~\ref{tab:systbig} for the \Jpsi~and \Psis,
respectively.

\section{Results}\label{sec:resu}

The prompt and non-prompt double differential cross sections for the two
charmonium states are obtained by multiplying the measured inclusive
cross sections with the fraction of prompt and non-\-prompt events, respectively.
In addition the cross-section ratio of the two charmonium states is
calculated.

Statistical uncertainties and contributions from the investigated sources to the total systematic uncertainties on these cross sections are summarized in Tables~\ref{tab:systsmall} and~\ref{tab:systbig}. The largest
uncertainties are due to the efficiency
correlations; FSR estimation has
a sizeable effect only in bins close to the edges of the acceptance.

\subsection{Prompt and non-prompt cross sections corrected for acceptance}
\label{sec:corrresu}

\begin{figure}[htpb!]
\centering
{
\includegraphics[width=7.5cm]{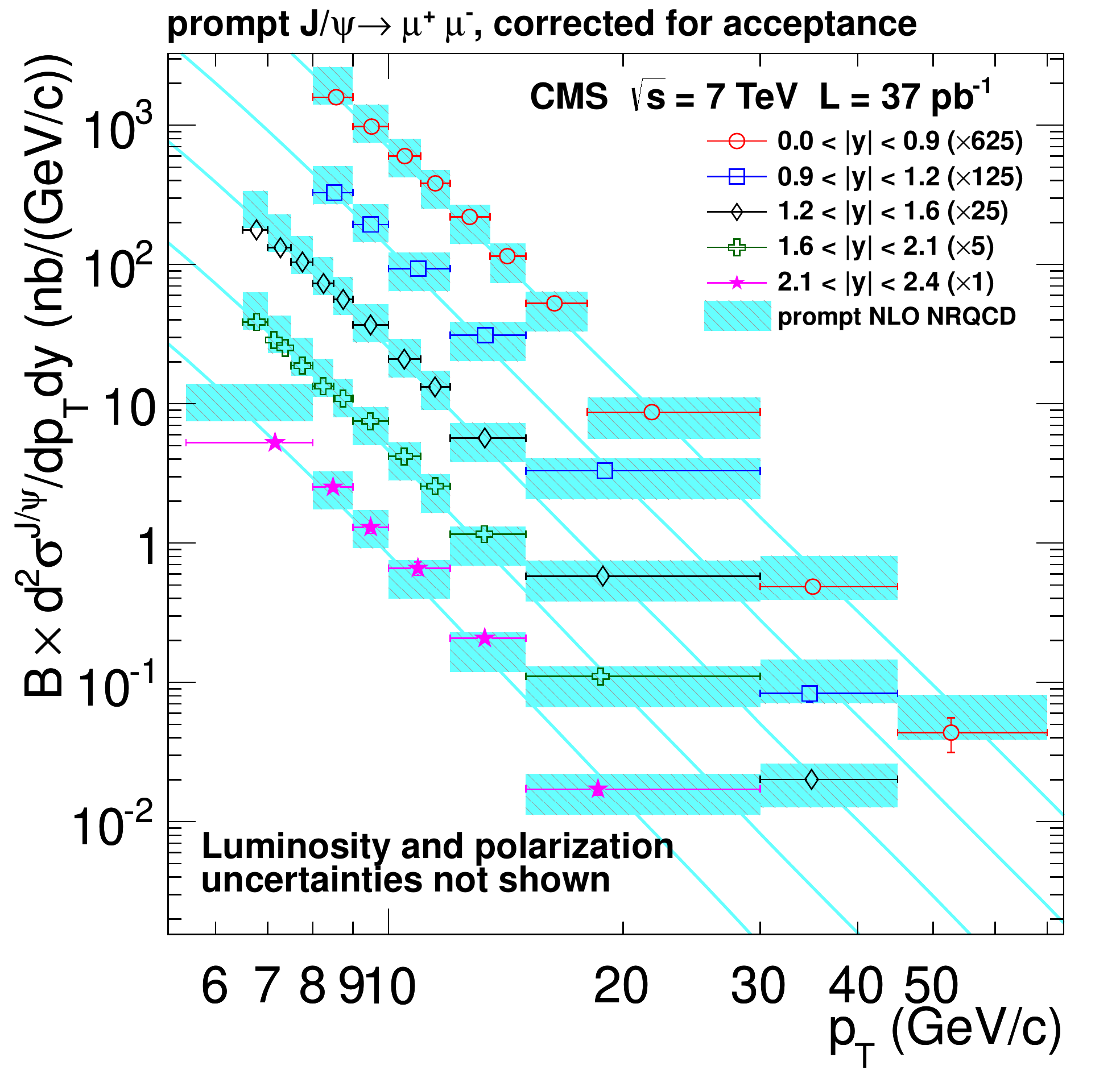}
\includegraphics[width=7.5cm]{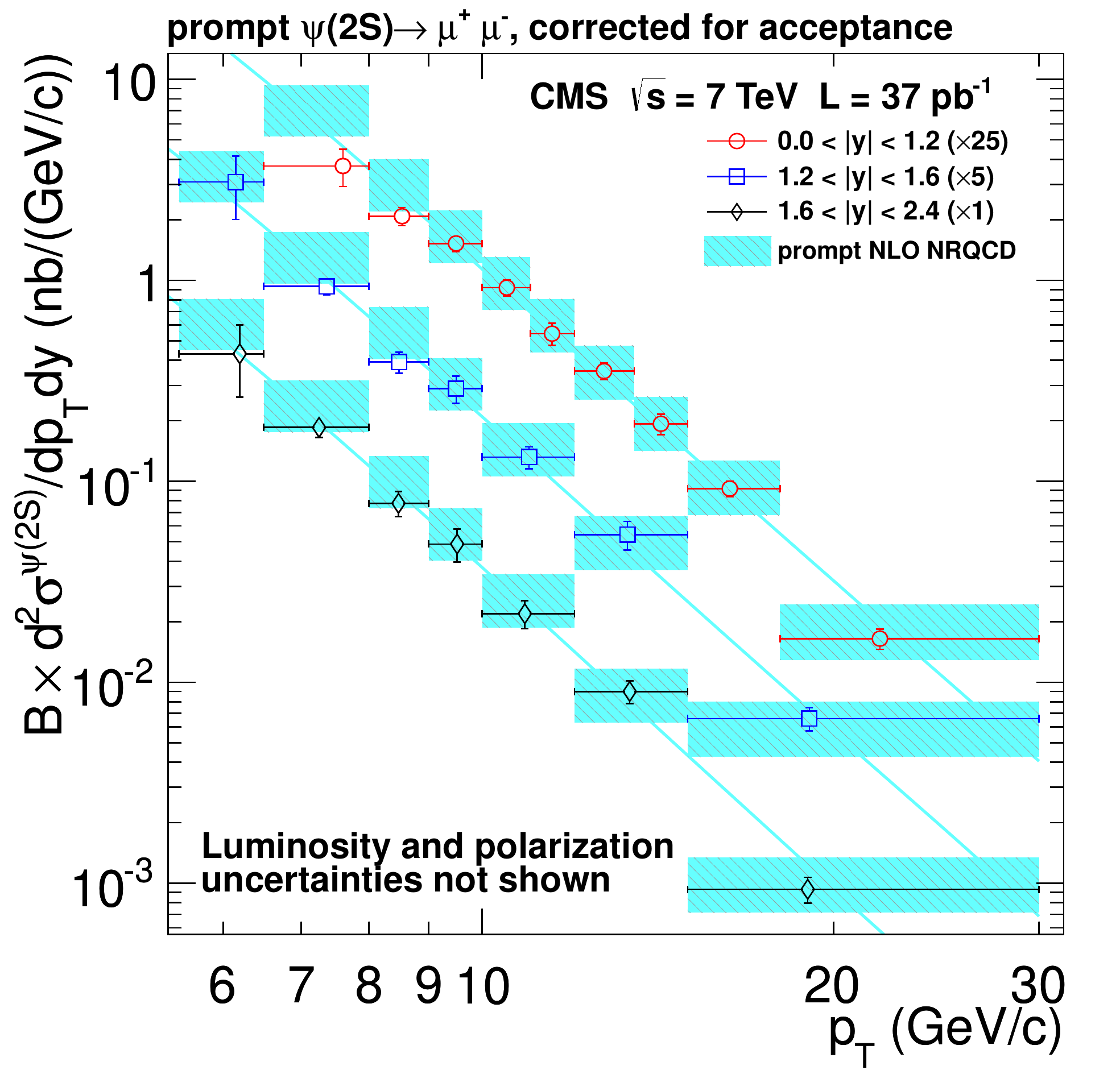}
}
\caption{Measured differential cross section for prompt \Jpsi\ and \Psis\ production
(left and right, respectively)
as a function of \mypt~for different rapidity bins.
The error bars on the data points include all the statistical and systematic contributions except luminosity and polarization.
The measurements have been offset by the numerical values given in the legend for easier viewing.
The coloured (dark) bands indicate the theoretical predictions from NRQCD calculations. The lines are added only for illustrative purposes.
}
\label{fig:XSprompt}
\end{figure}

\begin{figure}[htpb!]
\centering
{
\includegraphics[width=7.5cm]{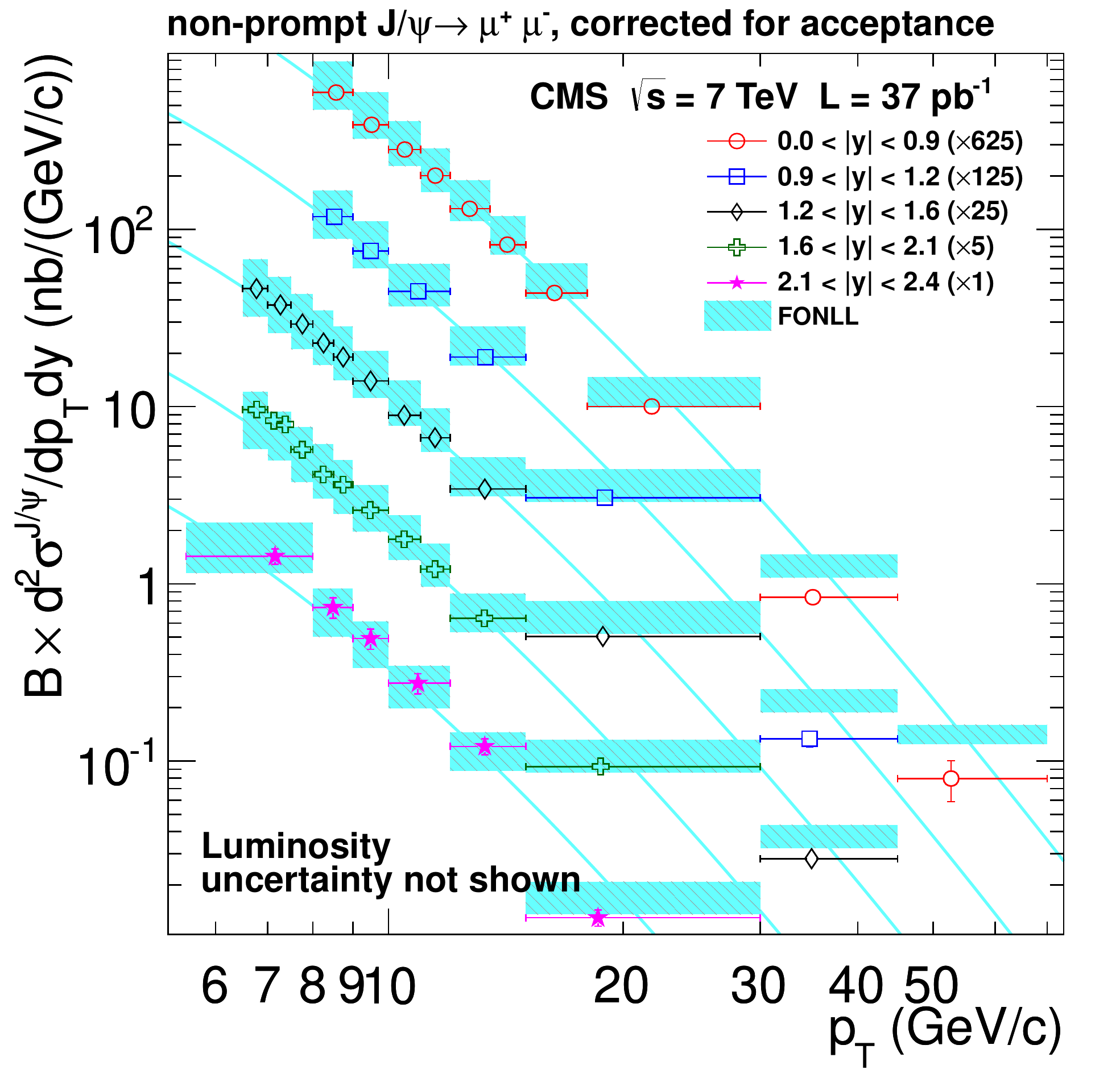}
\includegraphics[width=7.5cm]{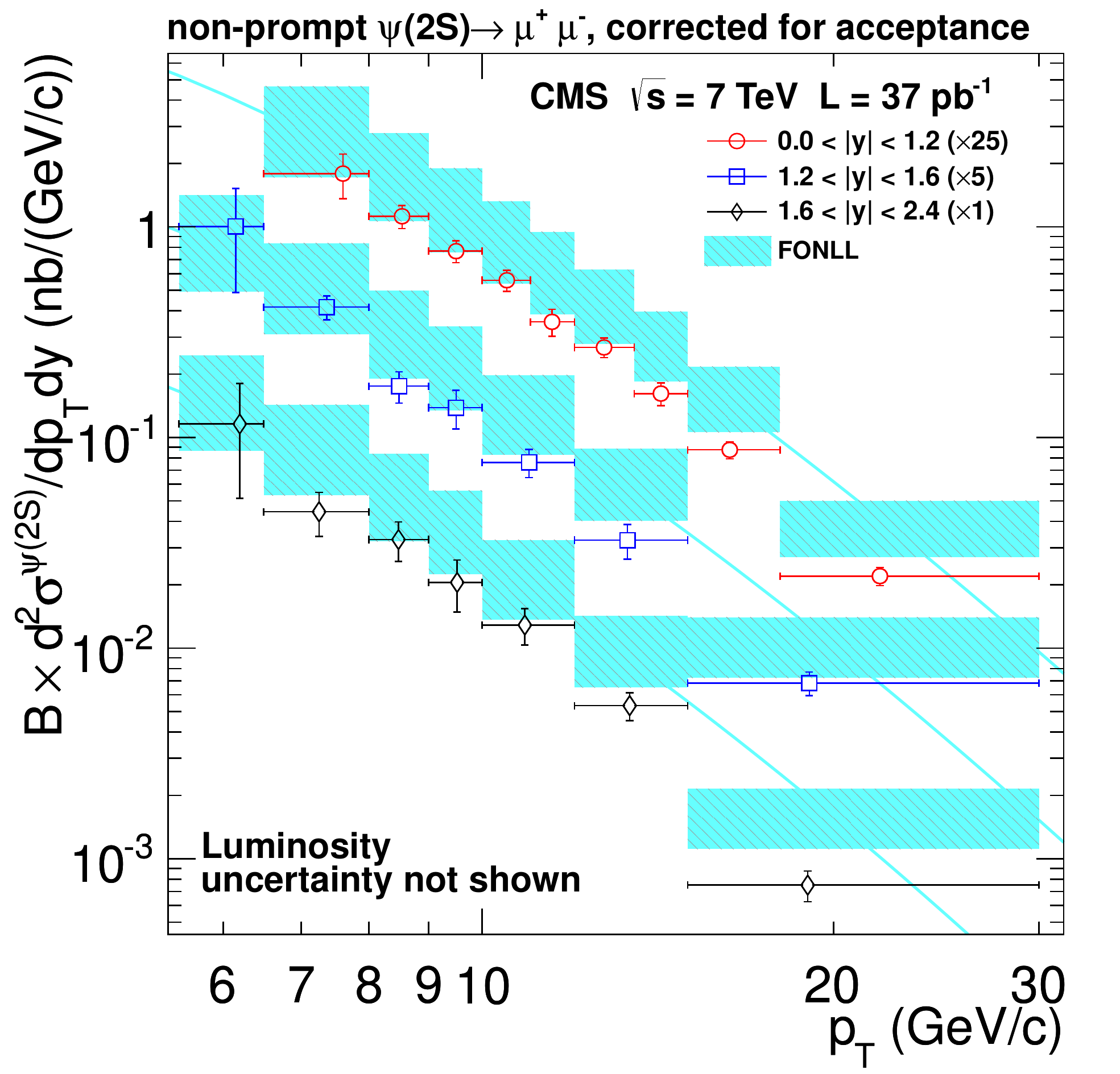}
}
\caption{Measured differential cross section for non-prompt \Jpsi\ and \Psis\
production (left and right, respectively)
as a function of \mypt~for different rapidity bins.
The error bars on the data points include all the statistical and systematic contributions except luminosity.
The measurements have been offset by the numerical values given in the legend for easier viewing.
The coloured (dark) bands indicate the theoretical predictions from FONLL calculations. The lines are added only for illustrative purposes.
}
\label{fig:XsecB}
\end{figure}

Figures~\ref{fig:XSprompt} and \ref{fig:XsecB} show the measured prompt and non-prompt cross sections
for the \Jpsi~and the \Psis\
as a function of \mypt, for the various rapidity bins and corrected for
detector acceptance. They are compared with theoretical predictions  from
NRQCD~\cite{Chao} and from  FONLL~\cite{cacciar1, cacciar2} for the prompt and non-prompt cases, respectively.
Numerical
values are also reported in Ref.~\cite{thetables}.

\begin{table}[tb]
\centering \caption{Summary of the relative
statistical and systematic uncertainties on the
non-prompt \Jpsi~cross section~(in \%). The variation  over the different \mypt~bins is given for the
five rapidity regions. Uncertainties on the prompt cross section are
identical, with the exception of the  non-prompt fraction, where they must be regarded
as relative to $(1-f_\cPqb)$ rather than to $f_\cPqb$.
Acceptance uncertainties on the FSR are given,
excluding the lowest-\mypt\ bin
in every rapidity region, where it can be as large as $19\%$ because of
acceptance edge effects.
}
\label{tab:systsmall}
\small
\begin{tabular}{llccccc}
\hline
$|y|$ range & & $0-0.9$ & $0.9-1.2$ & $1.2-1.6$ & $1.6-2.1$ & $2.1-2.4$ \\
\hline
Quantity & Source & \multicolumn{5}{c}{Relative uncertainty (in \%)} \\
affected & & & & & & \\
\hline \hline
\multicolumn{7}{c}{All cross sections}\\
\hline \hline
$m_{\mu \mu}$ fits & Statistical & $1.2-8.9$ & $1.5-7.1$ & $1.6-8.4$ & $1.2-3.2$ & $2.3-3.9$ \\
$\ell_{\Jpsi}$ fits & Statistical & $1.0-5.9$ & $1.4-4.7$ & $1.4-7.6$ & $2.1-8.3$ & $4.4-7.1$ \\
\hline
Efficiency & Single-muon efficiency & $0.3-0.9$ & $0.2-1.6$ & $0.1-1.4$ & $0.2-1.0$ & $0.6-1.4$ \\
 & $\rho$ factor & $1.9-23.2$ & $1.2-7.6$ & $0.7-5.7$ & $0.8-5.4$ & $3.7-6.8$ \\
Yields & Fit functions & $0.6-3.4$ & $0.4-2.8$ & $0.5-2.8$ & $0.8-2.2$ & $1.0-4.2$ \\
Luminosity & Luminosity & 4.0 & 4.0 & 4.0 & 4.0 & 4.0 \\
\hline
Non-prompt & Tracker misalignment & $0.1-2.1$ & $0.1-0.8$ & $0.0-1.5$ & $0.2-3.2$ & $0.2-5.1$ \\
fraction & b-lifetime model & $0.1-3.0$ & $0.1-3.4$ & $0.1-3.7$ & $0.2-2.6$ & $0.2-6.6$ \\
& Vertex estimation & $0.1-0.7$ & $0.7-3.0$ & $0.4-3.7$ & $1.5-4.6$ & $2.3-5.0$ \\
& Background fit &  $0.0-0.2$ & $0.1-1.4$ & $0.1-1.0$ & $0.0-2.5$ & $0.1-1.2$ \\
& Resolution model & $0.2-3.5$ & $0.0-4.2$ & $0.8-3.5$ & $1.1-5.0$ & $1.1-4.4$ \\
& Efficiency & $0.4-2.1$ & $0.9-3.3$ & $0.5-9.9$ & $0.3-3.3$ & $1.6-10.5$ \\
\hline \hline
\multicolumn{7}{c}{Only acceptance-corrected cross sections}\\
\hline \hline
Acceptance & FSR & $0.0-1.5$ & $0.0-2.5$ & $0.0-4.2$ & $0.7-8.0$ & $0.5-3.5$ \\
 & $\mypt$ calibration & $0.0-0.6$ & $0.0-0.6$ & $0.0-0.8$ & $0.1-0.6$ & $0.0-0.8$ \\
 & Kinematic spectra & $0.0-0.3$ & $0.0-0.7$ & $0.0-0.7$ & $0.7-3.8$ & $0.4-5.3$ \\
 & \PB\ polarization & $0.0-0.5$ & $0.0-0.4$ & $0.0-0.5$ & $0.1-0.8$ & $0.3-1.3$ \\
\hline
\end{tabular}
\end{table}

\begin{table}[bt]
\centering \caption{Summary of the relative
statistical and systematic uncertainties in the
non-prompt
\Psis~cross section~(in \%). The variation  over the different \mypt~bins is given for the
three rapidity regions.
Uncertainties on the prompt cross section are
identical, with the exception of the non-prompt fraction, where they must be regarded
as relative to $(1-f_\cPqb)$ rather than to $f_\cPqb$.
Acceptance uncertainties
on the FSR are given excluding the lowest-\mypt\ bin
in every rapidity region, where it can be as large as $29\%$ because of
acceptance edge effects.
}
\label{tab:systbig}
\small
\begin{tabular}{llccc}
\hline
$|y|$ range & & $0-1.2$ & $1.2-1.6$ & $1.6-2.4$ \\
\hline
Quantity & Source & \multicolumn{3}{c}{Relative uncertainty (in \%)} \\
affected & & & & \\
\hline \hline
\multicolumn{5}{c}{All cross sections}\\
\hline \hline
$m_{\mu \mu}$ fits & Statistical & $5.6-14.8$ & $7.5-31.7$ & $7.3-24.1$ \\
$\ell_{\Psis}$ fits & Statistical & $4.3-12.7$ & $5.9-38.0$ & $9.1-26.4$ \\
\hline
Efficiency & Single-muon efficiency & $0.1-0.5$ & $0.1-0.6$ & $0.2-0.9$ \\
 & $\rho$ factor & $0.7-13.1$ & $2.1-6.6$ & $2.3-9.8$ \\
Yields & Fit functions & $1.2-3.7$ & $0.6-12.1$ & $3.1-10.0$ \\
Luminosity & Luminosity & 4.0 & 4.0 & 4.0  \\
\hline
Non-prompt  & Tracker misalignment & $0.3-2.6$ & $1.5-7.1$ & $1.8-11.1$  \\
fraction & \cPqb-lifetime model & $0.0-2.5$ & $0.4-7.6$ & $0.0-2.9$ \\
& Vertex estimation & $0.0-1.7$ & $0.2-3.5$ & $1.2-4.2$  \\
& Background fit &  $1.0-6.8$ & $2.2-10.0$ & $2.5-15.3$ \\
& Resolution model & $0.5-3.5$ & $0.1-4.6$ & $0.9-24.9$  \\
& Efficiency & $0.5-7.8$ & $0.9-6.3$ & $0.5-13.8$  \\
\hline \hline
\multicolumn{5}{c}{Only acceptance-corrected cross sections}\\
\hline \hline
Acceptance & FSR & $0.0-3.9$ & $0.5-3.4$ & $0.3-4.1$ \\
 & $\mypt$ calibration & $0.2-0.5$ & $0.3-0.5$ & $0.3-0.5$ \\
 & Kinematic spectra & $0.1-1.2$ & $0.0-0.9$ & $0.7-2.0$ \\
 & B polarization & $0.1-0.8$ & $0.0-0.6$ & $0.2-1.7$ \\
\hline
\end{tabular}
\end{table}

The NRQCD prediction includes
non-prompt production
in the \Jpsi\ case
caused by feed-down decays from heavier charmonia,
and can therefore
be directly compared with the data.
Good agreement is found in both the \Jpsi\ and the \Psis\ cases.
For non-prompt production
the measurements lie systematically below the FONLL predictions,
possibly because of the large uncertainty on the $\PB \to \Psis X$
branching fraction. In general, for both states, the observed
differential cross sections seem
to fall more rapidly than the FONLL prediction at high \mypt , and this effect
is more evident for the \Jpsi\  because of the higher \mypt~reach.

The NRQCD theoretical uncertainties include those on the feed-down contributions and on the
colour-octet, long-distance matrix elements determined from fits to the Tevatron data.
The FONLL theoretical errors
include uncertainties on $\mathcal{B}(\PB \to \Jpsi\ X)$ and
$\mathcal{B}(\PB \to \Psis X)$,
renormalization and factorization scales, \cPqb-quark and \cPqc-quark
masses, parton distribution functions, and \cPqb fragmentation parameters.

However,
uncertainties on the $\PB \to$ charmonium decay spectrum were not included in
the original FONLL prediction. To estimate those, we make use of the
\textsc{EvtGen} MC generator, which describes $\PB \to$ charmonium decays using a
sum of many exclusive modes. We split the decay modes into two categories,
``high-$Q$'' and ``low-$Q$'', if the value of $Q$
 in the decay is respectively
greater than or less than 1.2\GeVcc, where $Q = m_\PB - \sum_i m_i$ and the
index $i$ runs over the \PB\ decay products. As low-$Q$ (high-$Q$) modes yield
charmonia with smaller (larger) momentum in the \PB rest frame, they
populate different regions of the \PB-decay spectrum.
Two sets of non-prompt charmonium
MC events are generated according to the following criteria. In the
first, each high-$Q$ mode branching fraction is increased by
its world-average uncertainty~\cite{bib-pdg} or
by 100\% of its value
if the branching fraction is not measured.
Low-$Q$ mode branching
fractions are decreased by a
similar amount, rescaling the sum to unity after this procedure.
In the second, the treatment of the high- and low-$Q$ modes is interchanged.
The maximum
difference in the resulting spectra in the two cases is added to the
theoretical FONLL uncertainty.

To investigate the effect of the assumed \Jpsi\ and \Psis\ polarizations
on the prompt cross section,
the acceptance is recomputed for four
extreme 
polarization scenarios
corresponding to fully longitudinal or fully transverse
polarization
in the helicity and Collins-Soper frames~\cite{bib-faccioli}.
This produces 
 relative cross-section shifts across the entire kinematic range
of up to 18--20\% (20--25\%) for the \Jpsi\ (\Psis ) in the helicity frame,
and 6--15\%  for both states 
in the Collins-Soper frame. Detailed results can be found
in  Ref.~\cite{thetables}.

\subsection{Prompt and non-prompt cross sections uncorrected for acceptance}
As discussed previously, since the polarization effects are large compared
to the measurement uncertainties, cross-section values are also reported
that are restricted to the CMS muon acceptance region, to allow future measurements of
the \Jpsi\ and \Psis~polarization
to be exploited.

\begin{figure}[htpb!]
\centering
{
\includegraphics[width=7.5cm]{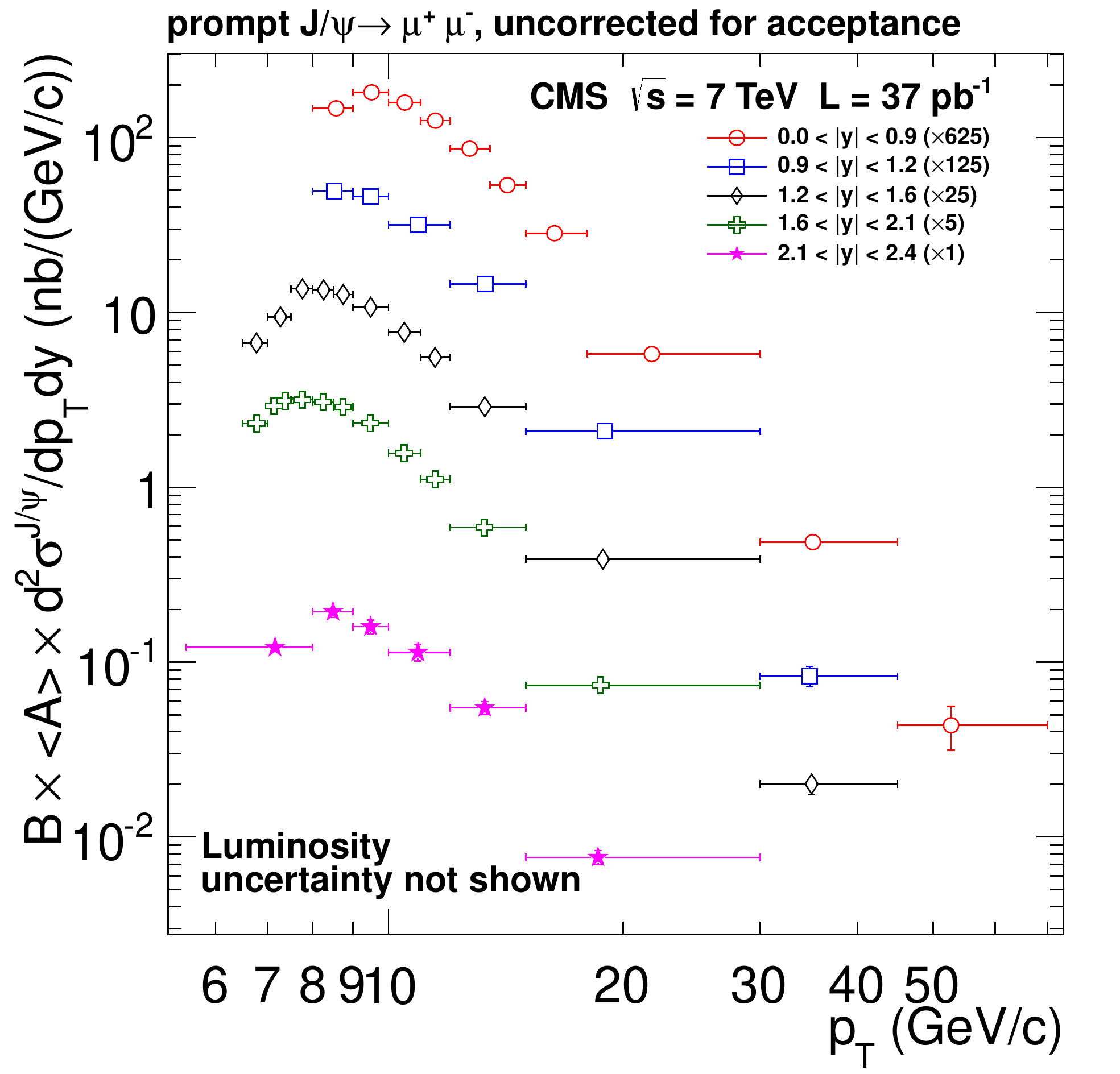}
\includegraphics[width=7.5cm]{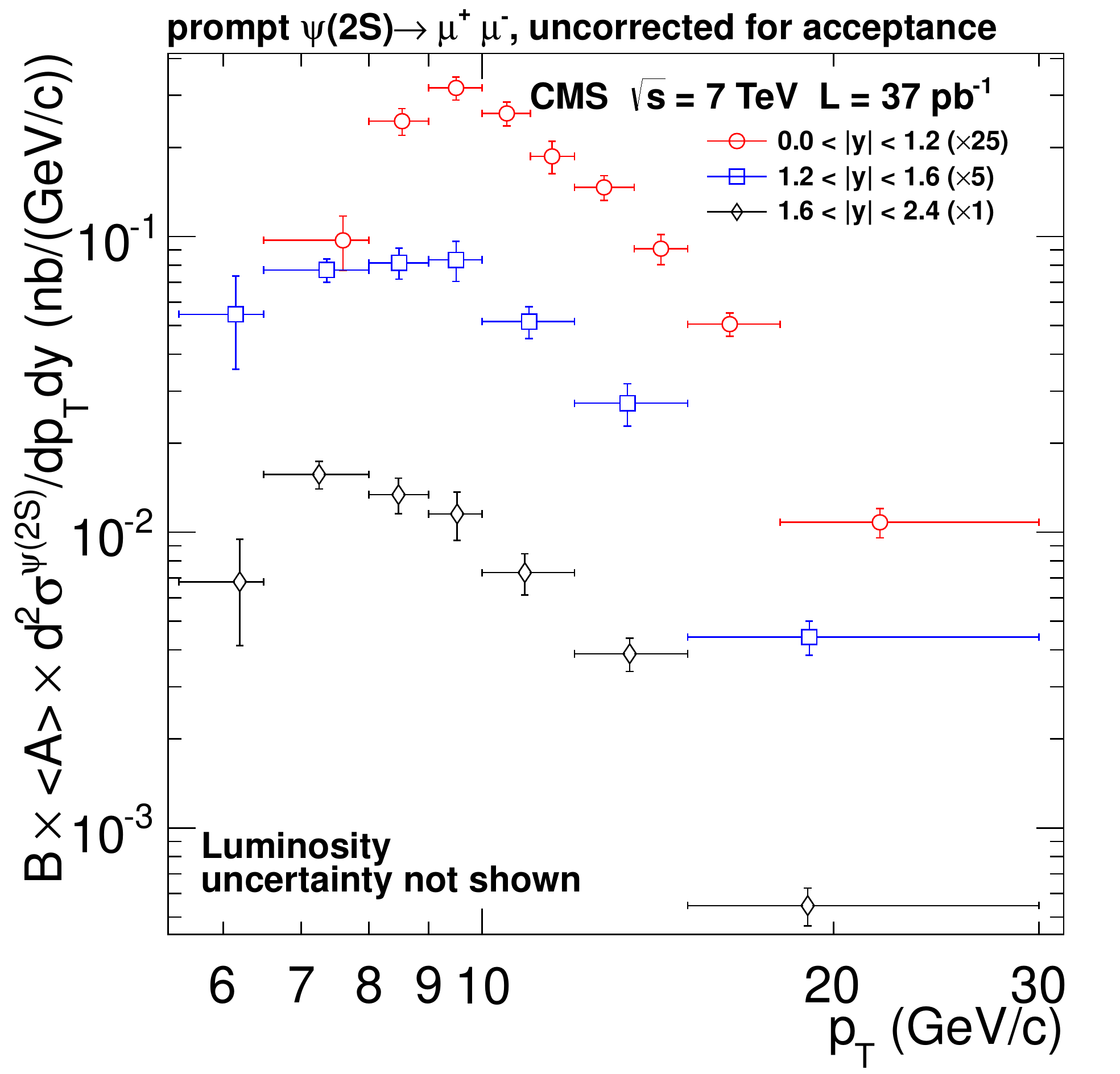}
}
\caption{Measured differential cross section for prompt \Jpsi~(left) and \Psis~(right) production
as a function of \mypt~for the different rapidity bins.
The error bars on data points include all the statistical and systematic contributions except luminosity.
The measurements have been offset by the numerical values given in the legend for easier viewing.
The results are not corrected for the muon acceptance.
}
\label{fig:XSpromptAcc}
\end{figure}

\begin{figure}[htpb!]
\centering
{
\includegraphics[width=7.5cm]{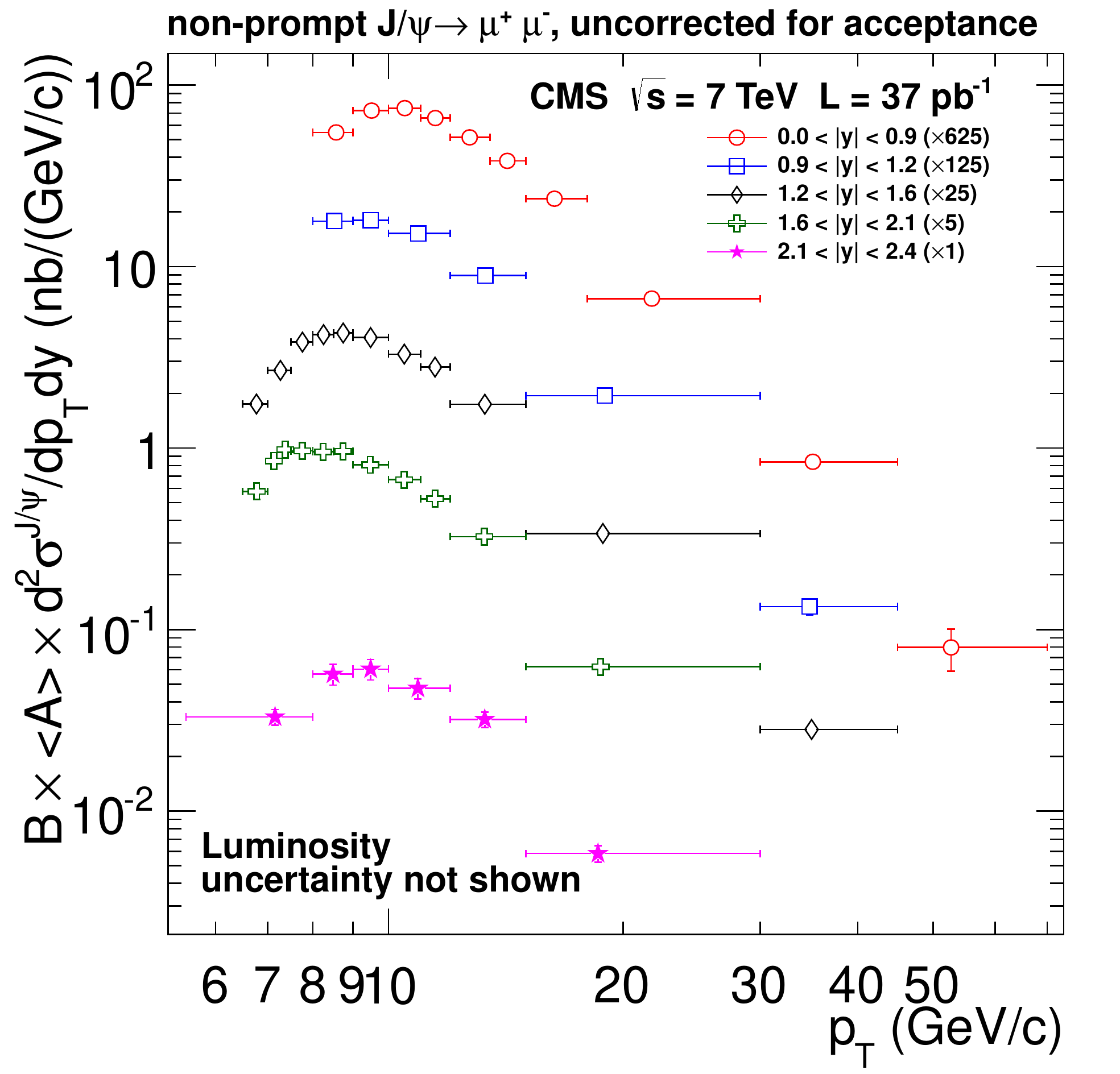}
\includegraphics[width=7.5cm]{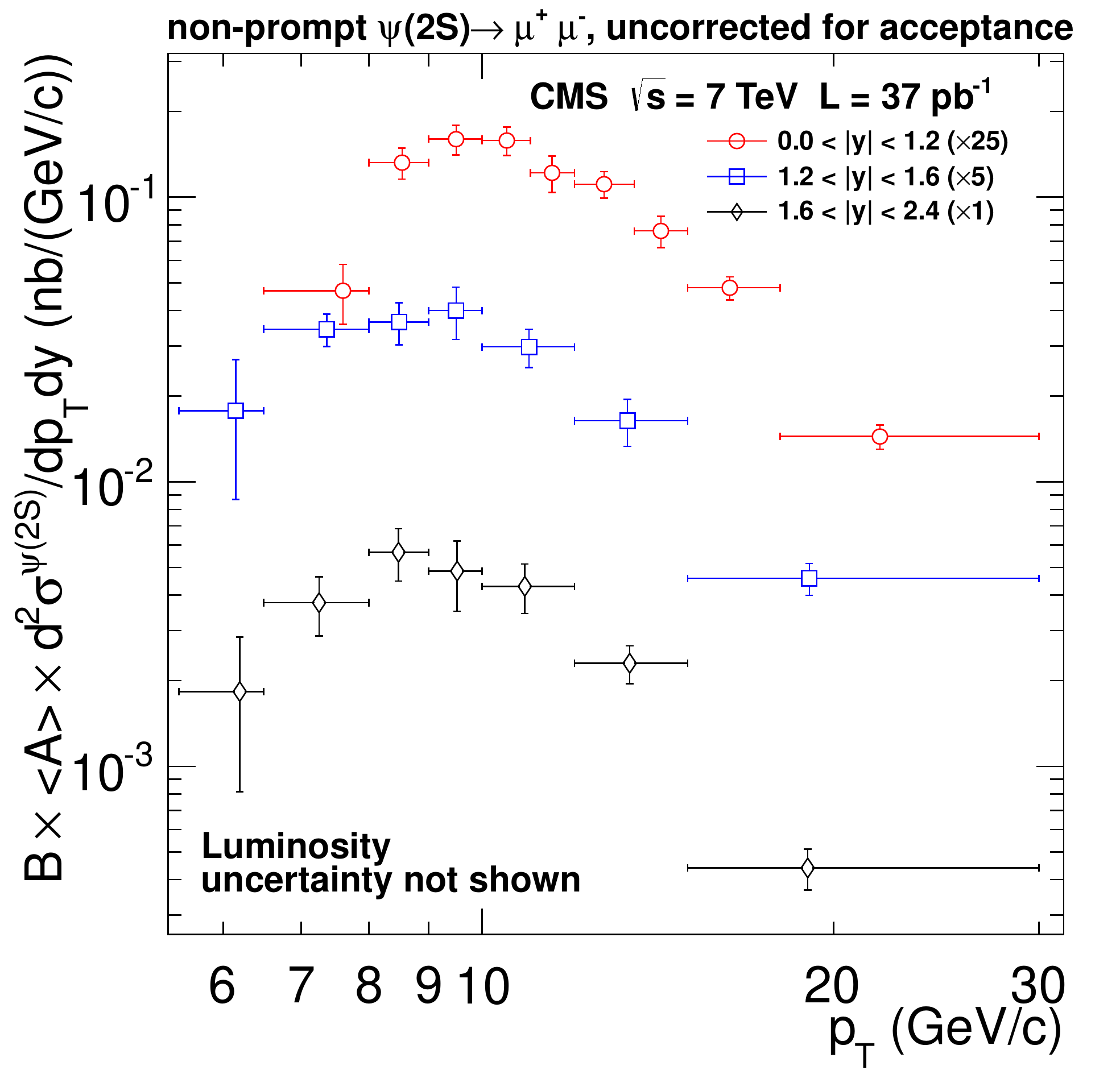}
}
\caption{Measured differential cross section for non-prompt \Jpsi~(left) and \Psis~(right) production
as a function of \mypt~for the different rapidity bins.
The error bars on data points include all the statistical and systematic contributions except luminosity.
The measurements have been offset by the numerical values given in the legend for easier viewing.
The result is not corrected for the  muon acceptance.
}
\label{fig:XsecBAcc}
\end{figure}

Figures~\ref{fig:XSpromptAcc} and \ref{fig:XsecBAcc} show the measured prompt and non-prompt cross sections
for the \Jpsi\ and the \Psis\
as a function of \mypt\ for the various rapidity bins and uncorrected for
detector acceptance. Numerical
values can be found in Ref.~\cite{thetables}.

\subsection{Non-prompt fractions}

\begin{figure}[tb]
\centering
{
\includegraphics[width=5cm,angle=90]{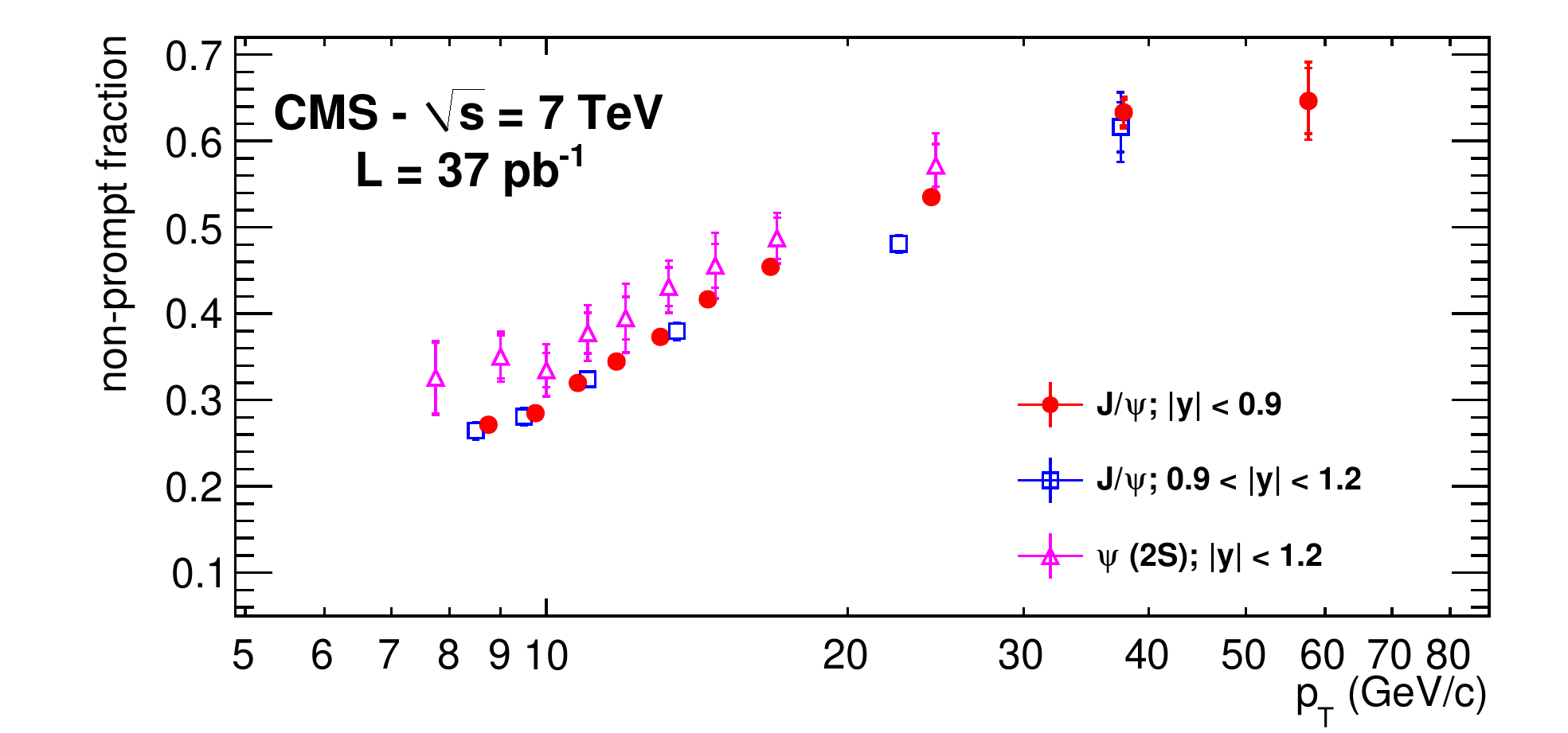} \\
\includegraphics[width=5cm,angle=90]{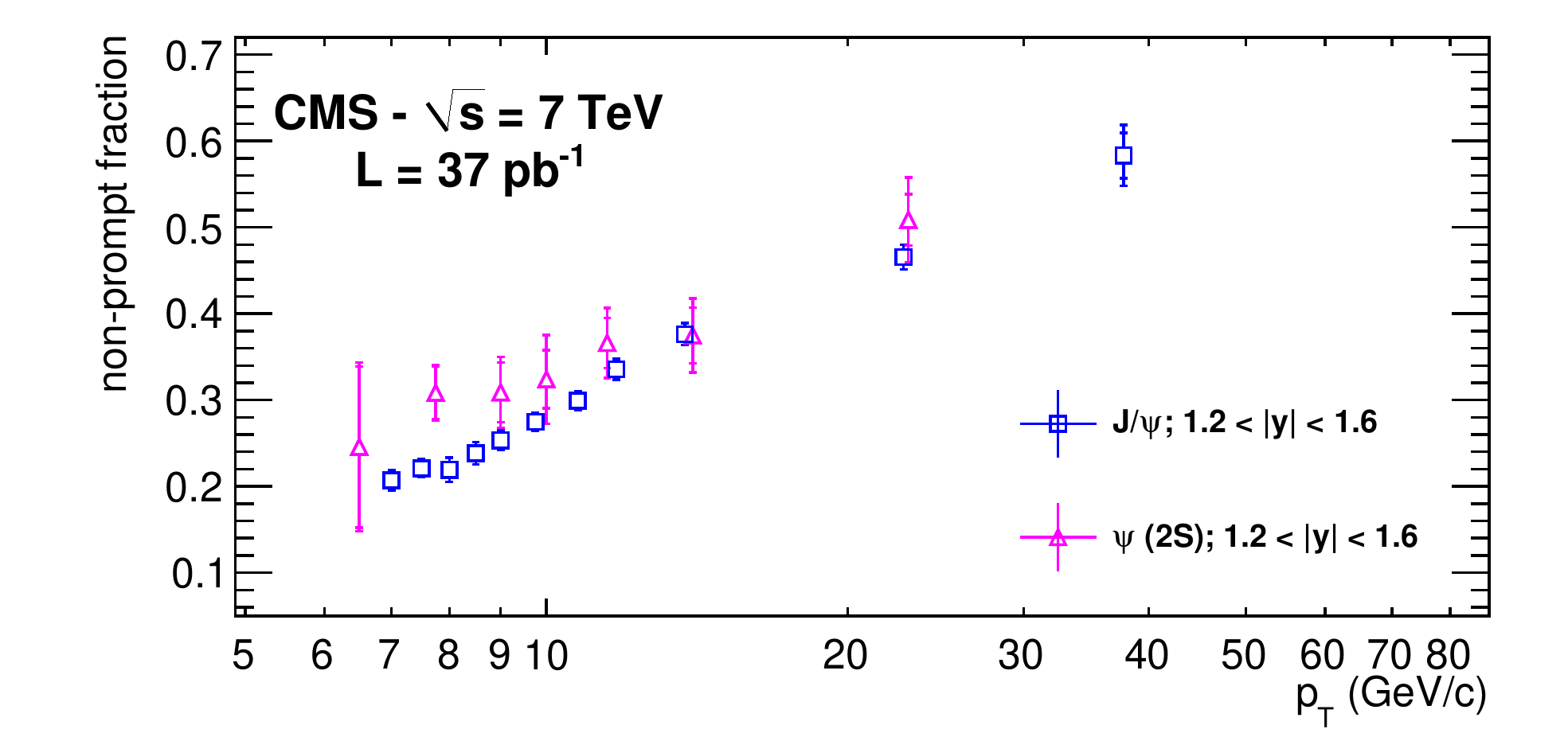} \\
\includegraphics[width=5cm,angle=90]{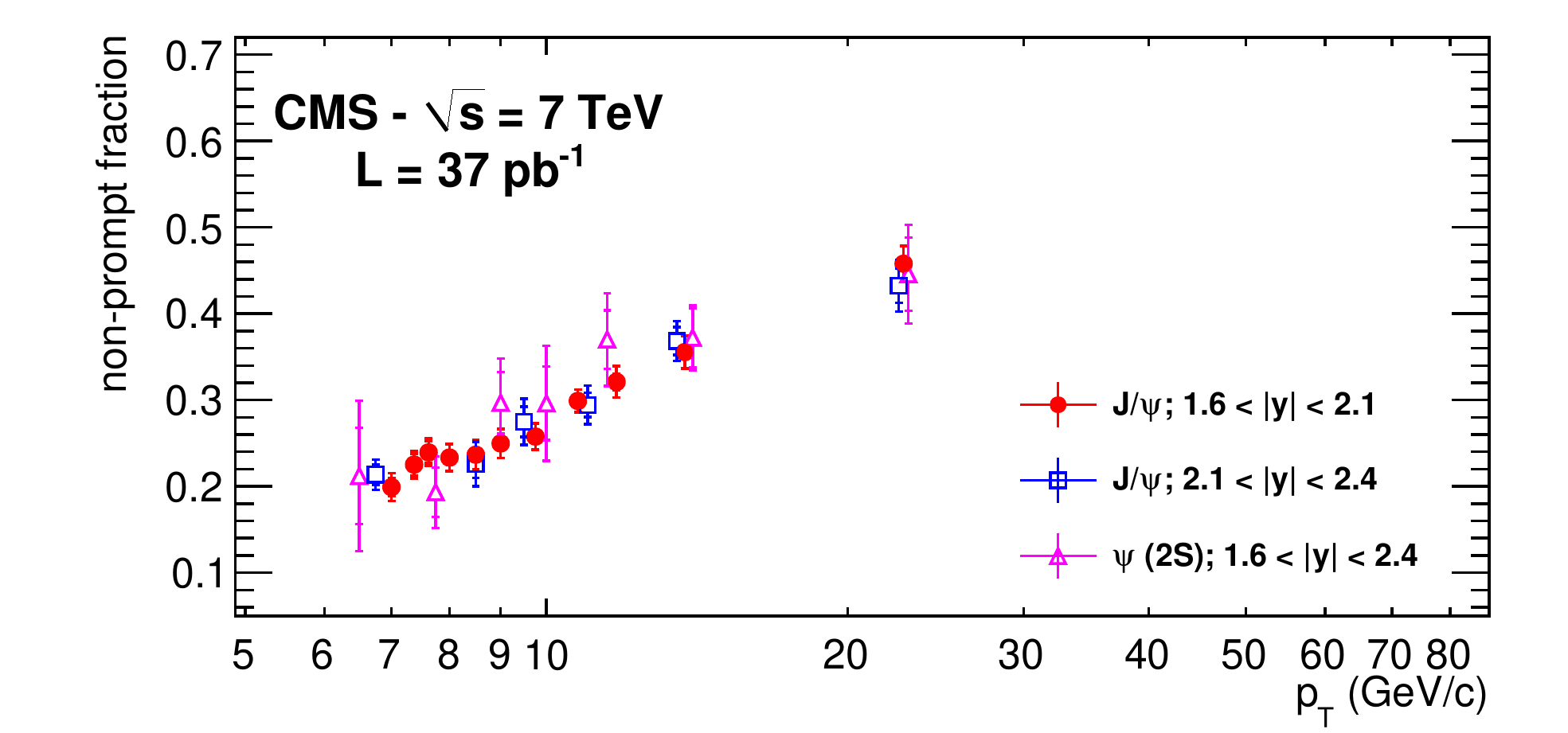}
}
\caption{Fitted \Jpsi~and \Psis~non-prompt fractions plotted as a function
of \mypt~for three rapidity regions: $0<|y|<1.2$ (top); $1.2<|y|<1.6$ (middle);
$1.6<|y|<2.4$ (bottom). The inner error bars represent the statistical
uncertainties only, the outer ones are the quadratic sum of statistical
and systematic uncertainties.}
\label{fig:bfractions}
\end{figure}

The measured non-prompt fractions for \Jpsi~and \Psis~mesons, extracted as described in Section~\ref{sec:nonprompt} and uncorrected for acceptance,
are reported in Ref.~\cite{thetables} and shown in Fig.~\ref{fig:bfractions}.
The uncertainties shown are statistical and systematic, and the measured
values are plotted as a function of \mypt~in three rapidity ranges. In
agreement with previous measurements~\cite{JPsiPaper,cdfxs_psip}, we observe
similar sizes of non-prompt fractions for \Jpsi~and \Psis , and
an increasing trend with \mypt.
Acceptance corrections do not induce significant changes in the non-prompt
fractions within their uncertainties.

\subsection{Cross-section ratio\label{sec:ratio}}

Most of the systematic uncertainties on
the
acceptances and
efficiencies listed in Tables~\ref{tab:systsmall} and~\ref{tab:systbig}, as well as the luminosity uncertainty, cancel
partially or fully
in the ratio of the \Psis\ to \Jpsi\ cross sections.
For this reason we also present the ratio of the two differential cross sections:

\begin{equation}
R(\mypt,|y|)=\frac{\frac{\mathrm{d}^2\sigma}{\mathrm{d}\mypt \mathrm{d}y}(\Psis) \cdot \mathcal{B}(\Psis\rightarrow \mu^{+}\mu^{-}) }{\frac{\mathrm{d}^2\sigma}{\mathrm{d}\mypt \mathrm{d}y}(\Jpsi) \cdot \mathcal{B}(\Jpsi\rightarrow \mu^{+}\mu^{-}) }=
\frac{N^{\text{corr}}_{\Psis}(\mypt , |y|)}{N^{\text{corr}}_{\Jpsi}(\mypt , |y|)} \quad ,
\label{eq:ratio}
\end{equation}
where the ratio $R$ is computed in bins of \mypt~and rapidity,
and the binning is
the same as used for
the \Psis~cross section.

The statistical uncertainties affecting $R$
are extracted directly from the simultaneous invariant mass fits.
The systematic uncertainties are estimated by
considering the same sources as for the cross sections (except the luminosity
and single-muon efficiency, which cancel out in the ratio)
and evaluating directly the variation of the ratio, in order to take correlations into account.

\begin{figure}[tbp]
\centering
{
\includegraphics[width=7.2cm]{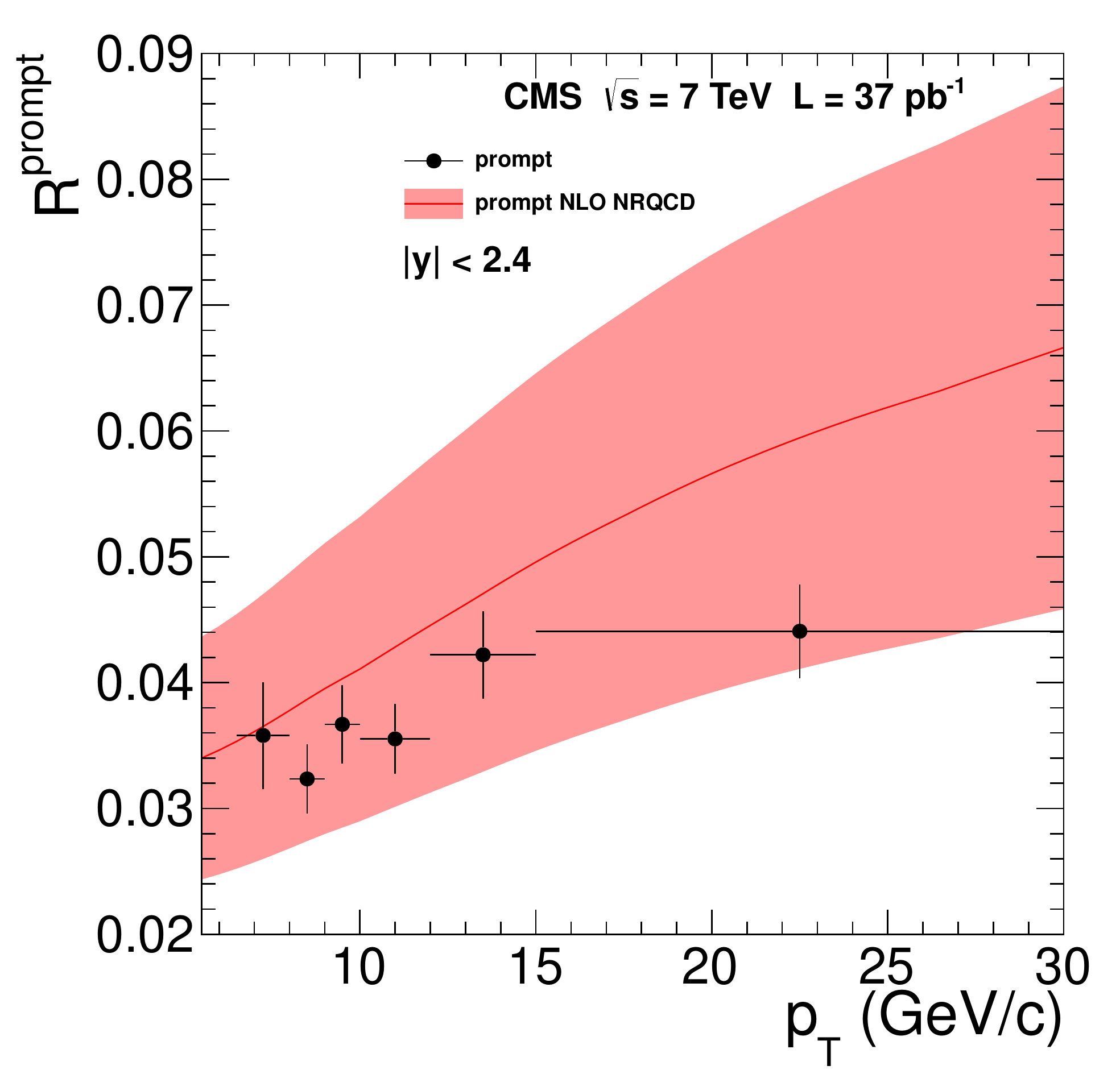}
\includegraphics[width=7.2cm]{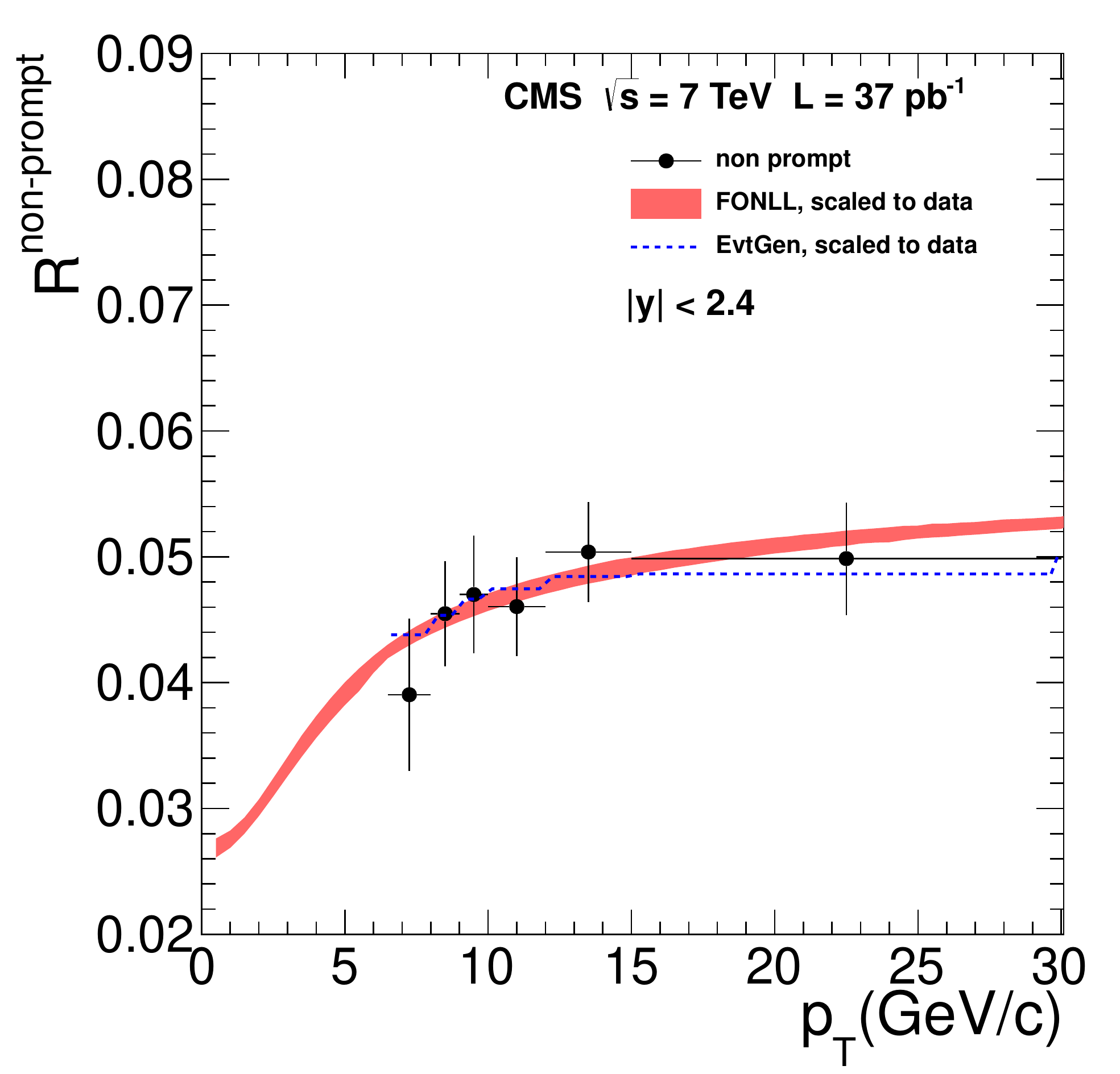}
}
\caption{
Measured value of $R$, the \Psis~to \Jpsi~differential cross-section ratio defined in Eq.~\ref{eq:ratio}, for prompt (left) and non-prompt (right) production, averaged over rapidity and plotted
as a function of \mypt. The left plot also includes the comparison
with the NRQCD prediction, while the right plot shows
the predictions of the
theoretical models used to determine
$\mathcal{B}(\PB \to \Psis X)$,
after the latter have been rescaled
to the fitted value given in Eq.~(\ref{eq:BR}).
The shaded bands show the uncertainties on the theoretical predictions.
The error bars give the total uncertainties on the
measurements; polarization
uncertainties are not included.}
\label{fig:XsRatio}
\end{figure}

No significant dependence of $R$ on
rapidity is observed; the  ratios  over the entire
rapidity range are therefore
computed. The resulting prompt and non-prompt cross-section ratios are shown
in Fig.~\ref{fig:XsRatio} as a function of \mypt . Numerical values of
rapidity-dependent and integrated ratios are given in Ref.~\cite{thetables}.

The assumptions on
polarization also affect the prompt  cross-section ratio
measurement.
In a plausible scenario~\cite{facciolipriv}, the
polarizations of the directly produced \Jpsi~and \Psis~states
are assumed to be the same. Therefore the uncertainty on the ratio comes only
from the difference between the polarization of the directly
produced mesons and the polarization of the \Jpsi~coming from decays of
P-wave states ($\chi_{c1}$ and $\chi_{c2}$), for which the maximum possible
variations are considered.
Using the measured feed-down fractions measured
at CDF~\cite{cdfchic1,cdfchic2},
this leads to the definition of the two extreme scenarios:
\begin{itemize}
\item $\lambda_\theta^{\Psis} = 1$, $\lambda_\theta^{\Jpsi} = 0.445$;
\item $\lambda_\theta^{\Psis} = -1$, $\lambda_\theta^{\Jpsi} = -0.647$;
\end{itemize}
which result in changes to the measured prompt cross-section ratio by $12-20\%$.

\subsection{Inclusive \texorpdfstring{$\PB \to \Psis X$}{B to psi(2S) X} branching fraction\label{sec:brpsi}}

The non-prompt \Psis~cross-section results can be used to determine
$\mathcal{B}(\PB \to \Psis X)$, the average inclusive branching fraction of
all weakly decaying particles containing
a \cPqb quark to \Psis .

Since the results are determined only for a limited range of phase
space, a theoretical assumption is needed to extrapolate to the full phase space.
The most precise result is obtained using the non-prompt
cross-section ratio, where most theoretical uncertainties cancel.
The FONLL model is used for the
result, taking as an alternative the
{\sc EvtGen} prediction
to determine
a systematic uncertainty.

For both models, the  predicted ratio is computed
for each \mypt~bin used in
the measurement, assuming the world-average values, listed by the Particle Data Group (PDG)~\cite{bib-pdg}, of
$\mathcal{B}_{\mathrm{PDG}}(\PB \to \Jpsi~X)$,
$\mathcal{B}_{\mathrm{PDG}}(\Psis \to \mu^+ \mu^-)$, and $\mathcal{B}_{\mathrm{PDG}}(\Jpsi\ \to \mu^+ \mu^-)$
(the branching fractions taken from the PDG are indicated
as $\mathcal{B}_{\mathrm{PDG}}$). In both models the production
cross section for each type of B meson or baryon
is weighted using the
values of the fractions $f(\mathrm{b} \to \PB^0)$, $f(\mathrm{b} \to \PB^\pm)$,
$f(\mathrm{b} \to \PB_s^0)$, and $f(\mathrm{b} \to \Lambda_{\mathrm{b}}^0)$,
taken from LEP and Tevatron measurements.
The predictions are then fitted to the data points,
leaving only the normalization ($N$) as a free parameter.
A good agreement in the shape of the \mypt\ distribution is found
for both models. The branching fraction
$\mathcal{B}(\PB \to \Psis X)$ is then
derived from the fitted normalization.

In addition to the fit uncertainty, including statistical and systematic
uncertainties on the single measurements ($\pm 3.8\%$), the following sources
of uncertainty are considered
(with the corresponding relative $\Delta \mathcal{B}$ uncertainty given in
parentheses):
\begin{itemize}
\item \emph{PDG branching fractions.} The uncertainties quoted by the PDG for
$\mathcal{B}(\PB \to \Jpsi~X)$, $\mathcal{B}(\Psis\ \to \mu^+ \mu^-)$,  and
$\mathcal{B}(\Jpsi\ \to \mu^+ \mu^-)$ are summed in quadrature ($13.5\%$).
\item \emph{Model assumptions.}
The \textsc{EvtGen} prediction is used for an alternative fit and
the difference with respect to the nominal $N$ value is taken as a systematic
uncertainty ($1.0\%$).
\item \emph{FONLL uncertainties.}
All uncertainties on the underlying \bbbar cross section, discussed in
Section~\ref{sec:corrresu}, are assumed to be fully correlated between
$\PB \to \Jpsi~X$ and $\PB \to \Psis X$ transitions.
Residual uncertainties affecting the cross-section ratio prediction
are used to perform alternative fits and the differences with respect to the
nominal value of $N$ are taken as systematic uncertainties ($1.5\%$).
\item \emph{$\PB \to$ charmonium spectrum theoretical uncertainties.}
The only source of theoretical uncertainty which is not correlated is the one
on the $\PB \to$ charmonium spectrum. We use the high-$Q$ and low-$Q$ method
to estimate this uncertainty as detailed in Section~\ref{sec:corrresu}. In
order to obtain an
upper limit on the uncertainty of this ratio, the high-$Q$ sample
of $\PB \to \Jpsi X$ is compared to the low-$Q$ $\PB \to \Psis X$
sample and vice versa. The average difference in $N$ with respect to
the nominal \textsc{EvtGen} prediction is taken as a systematic uncertainty ($3.8\%$).

\end{itemize}

The measured value is:
\begin{equation}
\mathcal{B}(\PB \to \Psis X) = (3.08 \pm 0.12 (\text{stat.+syst.}) \pm 0.13 (\text{theor.}) \pm 0.42 (\mathcal{B}_\text{PDG})) \cdot 10^{-3} ,
\label{eq:BR}
\end{equation}
where the last uncertainties are from the world-average branching fractions and the theoretical variations, respectively. 
The result is in agreement with
the current world-average value from
LEP and Tevatron measurements
($\mathcal{B}_{\mathrm{PDG}}(\PB \to \Psis X) = (4.8 \pm 2.4) \cdot 10^{-3}$~\cite{bib-pdg}),
while improving the relative uncertainty by a factor of three.

\section{Summary\label{sec:conclusion}}
A measurement of the \Jpsi\ and
\Psis\
production cross sections
in $\Pp\Pp$ collisions at $\sqrt{s}=7\TeV$  with the CMS experiment at the LHC has
been presented.
The data sample corresponds to an integrated luminosity of
$36.7 \pm 1.5\pbinv$.
The two cross sections and their ratio
have been measured as a function of the meson transverse
momentum (up to 70\GeVc for the \Jpsi~and to 30\GeVc for the \Psis)
in several rapidity ranges.
Cross sections for prompt and non-prompt production have been determined
from the
measured values of the \Jpsi~and \Psis~non-prompt fractions.

The prompt cross section results are evaluated
assuming isotropic decays in the production, as well as four
other polarization scenarios.
In addition, cross sections restricted to the acceptance of the CMS
detector are given, which are not affected by the polarization of the
charmonium states.

Cross sections for prompt and non-prompt production
have been compared with NRQCD and FONLL predictions, respectively.
Agreement is found in the prompt case: this is particularly remarkable in the
\Psis~case, where theoretical uncertainties are reduced because of the absence of
feed-down from heavier charmonium states. In the non-prompt case,
general
agreement in shape is found for \Psis~in the entire \mypt~range considered (up
to 30\GeVc), but an overall
scale discrepancy is observed,
possibly because of
the assumption on the
inclusive branching fraction $\mathcal{B}(\PB \to \Psis X)$. For \Jpsi~there
is similarly general agreement over the above range, while the predictions overestimate
the measured differential cross-sections for $30 <  \mypt < 70$ GeV/$c$.

For plausible
hypotheses on the polarizations of the two charmonium states the ratio
of their differential cross sections is obtained. In this ratio
systematic errors largely cancel.
The inclusive branching fraction $\mathcal{B}(\PB \to \Psis X)$ is extracted from the  ratio
of the non-prompt cross sections to be:
$$
\mathcal{B}(\PB \to \Psis X) = (3.08 \pm 0.12\,(\text{stat.+syst.}) \pm 0.13\,(\text{theor.}) \pm 0.42 \,(\mathcal{B}_\text{PDG})) \times 10^{-3},
$$
improving the relative uncertainty on the
previous world average
by a factor of three.

\textbf{Acknowledgements}\\
\\
We would like to thank Yan-Qing Ma for providing theoretical predictions in
NLO NRQCD
and Matteo Cacciari for predictions in the FONLL scheme and
useful discussions.

\hyphenation{Bundes-ministerium Forschungs-gemeinschaft Forschungs-zentren} We wish to congratulate our colleagues in the CERN accelerator departments for the excellent performance of the LHC machine. We thank the technical and administrative staff at CERN and other CMS institutes. This work was supported by the Austrian Federal Ministry of Science and Research; the Belgium Fonds de la Recherche Scientifique, and Fonds voor Wetenschappelijk Onderzoek; the Brazilian Funding Agencies (CNPq, CAPES, FAPERJ, and FAPESP); the Bulgarian Ministry of Education and Science; CERN; the Chinese Academy of Sciences, Ministry of Science and Technology, and National Natural Science Foundation of China; the Colombian Funding Agency (COLCIENCIAS); the Croatian Ministry of Science, Education and Sport; the Research Promotion Foundation, Cyprus; the Estonian Academy of Sciences and NICPB; the Academy of Finland, Finnish Ministry of Education and Culture, and Helsinki Institute of Physics; the Institut National de Physique Nucl\'eaire et de Physique des Particules~/~CNRS, and Commissariat \`a l'\'Energie Atomique et aux \'Energies Alternatives~/~CEA, France; the Bundesministerium f\"ur Bildung und Forschung, Deutsche Forschungsgemeinschaft, and Helmholtz-Gemeinschaft Deutscher Forschungszentren, Germany; the General Secretariat for Research and Technology, Greece; the National Scientific Research Foundation, and National Office for Research and Technology, Hungary; the Department of Atomic Energy and the Department of Science and Technology, India; the Institute for Studies in Theoretical Physics and Mathematics, Iran; the Science Foundation, Ireland; the Istituto Nazionale di Fisica Nucleare, Italy; the Korean Ministry of Education, Science and Technology and the World Class University program of NRF, Korea; the Lithuanian Academy of Sciences; the Mexican Funding Agencies (CINVESTAV, CONACYT, SEP, and UASLP-FAI); the Ministry of Science and Innovation, New Zealand; the Pakistan Atomic Energy Commission; the State Commission for Scientific Research, Poland; the Funda\c{c}\~ao para a Ci\^encia e a Tecnologia, Portugal; JINR (Armenia, Belarus, Georgia, Ukraine, Uzbekistan); the Ministry of Science and Technologies of the Russian Federation, the Russian Ministry of Atomic Energy and the Russian Foundation for Basic Research; the Ministry of Science and Technological Development of Serbia; the Ministerio de Ciencia e Innovaci\'on, and Programa Consolider-Ingenio 2010, Spain; the Swiss Funding Agencies (ETH Board, ETH Zurich, PSI, SNF, UniZH, Canton Zurich, and SER); the National Science Council, Taipei; the Scientific and Technical Research Council of Turkey, and Turkish Atomic Energy Authority; the Science and Technology Facilities Council, UK; the US Department of Energy, and the US National Science Foundation.

Individuals have received support from the Marie-Curie programme and the European Research Council (European Union); the Leventis Foundation; the A. P. Sloan Foundation; the Alexander von Humboldt Foundation; the Belgian Federal Science Policy Office; the Fonds pour la Formation \`a la Recherche dans l'Industrie et dans l'Agriculture (FRIA-Belgium); the Agentschap voor Innovatie door Wetenschap en Technologie (IWT-Belgium); and the Council of Science and Industrial Research, India.

\newpage
\clearpage

\bibliography{auto_generated}   

\providecommand{\href}[2]{#2}\begingroup\raggedright\begin{thebibliography}{10}%
\makeatletter
\providecommand{\hrefCMSnoop }[0]{\@secondoftwo}%
\makeatother

\bibitem{Artoisenet:2007xi}
\hrefCMSnoop {} {P.~Artoisenet, J.~P. Lansberg, and F.~Maltoni,
  ``Hadroproduction of {$J/\psi$} and {$\Upsilon$} in association with a
  heavy-quark pair'',} \textit{ Phys. Lett. B} \textbf{ 653} (2007) 60.
  \href{http://dx.doi.org/10.1016/j.physletb.2007.04.031}{\texttt{
  doi:10.1016/j.physletb.2007.04.031}}.

\bibitem{Artoisenet:2007qm}
\hrefCMSnoop {} {P.~Artoisenet, F.~Maltoni, and T.~Stelzer, ``Automatic
  generation of quarkonium amplitudes in {NRQCD}'',} \textit{ JHEP} \textbf{
  02} (2008) 102.
  \href{http://dx.doi.org/10.1088/1126-6708/2008/02/102}{\texttt{
  doi:10.1088/1126-6708/2008/02/102}}.

\bibitem{Chao}
\hrefCMSnoop {} {Y.-Q. Ma, K.~Wang, and K.-T. Chao, ``{$\Jpsi (\psi ')$
  production at the Tevatron and LHC at $O(\alpha_s^4v^4)$ in nonrelativistic
  QCD}'',} \textit{ Phys. Rev. Lett.} \textbf{ 106} (2011) 042002,
  \href{http://www.arXiv.org/abs/1009.3655}{\texttt{ arXiv:1009.3655}}.
  \href{http://dx.doi.org/10.1103/PhysRevLett.106.042002}{\texttt{
  doi:10.1103/PhysRevLett.106.042002}}.

\bibitem{cdfxs_psip}
\hrefCMSnoop {} {{ CDF} Collaboration, ``Production of $\psi$(2S) mesons in
  $p{\bar p}$ collisions at 1.96 TeV'',} \textit{ Phys. Rev. D} \textbf{ 80}
  (2009) 031103. \href{http://dx.doi.org/10.1103/PhysRevD.80.031103}{\texttt{
  doi:10.1103/PhysRevD.80.031103}}.

\bibitem{cacciar1}
\hrefCMSnoop {} {M.~Cacciari, M.~Greco, and P.~Nason, ``The $p_{\mathrm{T}}$
  spectrum in heavy-flavour hadroproduction'',} \textit{ JHEP} \textbf{ 05}
  (1998) 007. \href{http://dx.doi.org/10.1088/1126-6708/1998/05/007}{\texttt{
  doi:10.1088/1126-6708/1998/05/007}}.

\bibitem{cacciar2}
\hrefCMSnoop {} {M.~Cacciari, S.~Frixione, and P.~Nason, ``The $p_{\mathrm{T}}$
  spectrum in heavy-flavour photoproduction'',} \textit{ JHEP} \textbf{ 03}
  (2001) 006. \href{http://dx.doi.org/10.1088/1126-6708/2001/03/006}{\texttt{
  doi:10.1088/1126-6708/2001/03/006}}.

\bibitem{JPsiPaper}
\hrefCMSnoop {} {{ CMS} Collaboration, ``Prompt and non-prompt {\Jpsi} cross
  sections in pp collisions at $\sqrt{s} = 7$~{TeV}'',} \textit{ Eur. Phys. J.
  C} \textbf{ 71} (2011) 1575.
  \href{http://dx.doi.org/10.1140/epjc/s10052-011-1575-8}{\texttt{
  doi:10.1140/epjc/s10052-011-1575-8}}.

\bibitem{atlas}
\hrefCMSnoop {} {{ ATLAS} Collaboration, ``Measurement of the differential
  cross-sections of inclusive, prompt and non-prompt {\Jpsi} production in
  proton-proton collisions at $\sqrt{s} = 7$~{TeV}'',} \textit{ Nucl. Phys. B}
  \textbf{ 850} (2011) 442.
  \href{http://dx.doi.org/10.1016/j.nuclphysb.2011.05.015}{\texttt{
  doi:10.1016/j.nuclphysb.2011.05.015}}.

\bibitem{lhcb}
\hrefCMSnoop {} {{ LHCb} Collaboration, ``Measurement of {\Jpsi} production in
  pp collisions at $\sqrt{s} = 7$~{TeV}'',} \textit{ Eur. Phys. J. C} \textbf{
  71} (2011) 1645.
  \href{http://dx.doi.org/10.1140/epjc/s10052-011-1645-y}{\texttt{
  doi:10.1140/epjc/s10052-011-1645-y}}.

\bibitem{alice}
\hrefCMSnoop {} {{ ALICE} Collaboration, ``Rapidity and transverse momentum
  dependence of inclusive {\Jpsi} production in pp collisions at $\sqrt{s} =
  7$~{TeV}'',} \textit{ Phys. Lett. B} \textbf{ 704} (2011) 442.
  \href{http://dx.doi.org/10.1016/j.physletb.2011.09.054}{\texttt{
  doi:10.1016/j.physletb.2011.09.054}}.

\bibitem{bib-evtgen}
\hrefCMSnoop {} {D.~J. Lange, ``The \textsc{EvtGen} particle decay simulation
  package'',} \textit{ Nucl. Instrum. Meth. A} \textbf{ 462} (2001) 152.
  \href{http://dx.doi.org/10.1016/S0168-9002(01)00089-4}{\texttt{
  doi:10.1016/S0168-9002(01)00089-4}}.

\bibitem{babarpol}
\hrefCMSnoop {} {{ BaBar} Collaboration, ``Study of inclusive production of
  charmonium mesons in B decays'',} \textit{ Phys. Rev. D} \textbf{ 67} (2003)
  032002. \href{http://dx.doi.org/10.1103/PhysRevD.67.032002}{\texttt{
  doi:10.1103/PhysRevD.67.032002}}.

\bibitem{JINST}
\hrefCMSnoop {} {{ CMS} Collaboration, ``The CMS experiment at the CERN LHC'',}
  \textit{ JINST} \textbf{ 03} (2008) S08004.
  \href{http://dx.doi.org/10.1088/1748-0221/3/08/S08004}{\texttt{
  doi:10.1088/1748-0221/3/08/S08004}}.

\bibitem{bib-lumi}
\href {http://cdsweb.cern.ch/record/1335668} {{ CMS} Collaboration, ``Absolute
  luminosity normalization'',} CMS Detector Performance Summary
  CERN-CMS-DP-2011-002, (2011).

\bibitem{bib-PYTHIA}
\hrefCMSnoop {} {T.~Sj{\"o}strand, S.~Mrenna, and P.~Z. Skands, ``\PYTHIA\ 6.4
  physics and manual'',} \textit{ JHEP} \textbf{ 05} (2006) 026.
  \href{http://dx.doi.org/10.1088/1126-6708/2006/05/026}{\texttt{
  doi:10.1088/1126-6708/2006/05/026}}.

\bibitem{kramer}
\hrefCMSnoop {} {M.~Kr{\"a}mer, ``Quarkonium production at high-energy
  colliders'',} \textit{ Prog. Part. Nucl. Phys.} \textbf{ 47} (2001) 141.
  \href{http://dx.doi.org/10.1016/S0146-6410(01)00154-5}{\texttt{
  doi:10.1016/S0146-6410(01)00154-5}}.

\bibitem{marianne}
\href {http://cdsweb.cern.ch/record/1042611} {M.~Bargiotti and V.~Vagnoni,
  ``Heavy quarkonia sector in {PYTHIA} 6.324: tuning, validation and
  perspectives at {LHC(b)}'',} CERN report LHCb-2007-042, (2007).

\bibitem{bib-photos1}
\hrefCMSnoop {} {E.~Barberio, B.~van Eijk, and Z.~W{\c{a}}s, ``\PHOTOS\ - a
  universal Monte Carlo for {QED} radiative corrections in decays'',} \textit{
  Comput. Phys. Commun.} \textbf{ 66} (1991) 115.
  \href{http://dx.doi.org/10.1016/0010-4655(91)90012-A}{\texttt{
  doi:10.1016/0010-4655(91)90012-A}}.

\bibitem{bib-photos2}
\hrefCMSnoop {} {E.~Barberio and Z.~W{\c{a}}s, ``\PHOTOS\ - a universal Monte
  Carlo for {QED} radiative corrections: version 2.0'',} \textit{ Comput. Phys.
  Commun.} \textbf{ 79} (1994) 291.
  \href{http://dx.doi.org/10.1016/0010-4655(94)90074-4}{\texttt{
  doi:10.1016/0010-4655(94)90074-4}}.

\bibitem{bib-magneticfield}
\hrefCMSnoop {} {{ CMS} Collaboration, ``{Precise mapping of the magnetic field
  in the CMS barrel yoke using cosmic rays}'',} \textit{ JINST} \textbf{ 05}
  (2010) T03021.
  \href{http://dx.doi.org/10.1088/1748-0221/5/03/T03021}{\texttt{
  doi:10.1088/1748-0221/5/03/T03021}}.

\bibitem{bib-material}
\href {http://cdsweb.cern.ch/record/1279138} {{ CMS} Collaboration, ``Studies
  of Tracker Material in the {CMS} Detector'',} CMS Physics Analysis Summary
  CMS-PAS-TRK-10-003, (2010).

\bibitem{bib-trackeralignment}
\hrefCMSnoop {} {{ CMS} Collaboration, ``{Alignment of the CMS silicon tracker
  during commissioning with cosmic rays}'',} \textit{ JINST} \textbf{ 5} (2010)
  T03009. \href{http://dx.doi.org/10.1088/1748-0221/5/03/T03009}{\texttt{
  doi:10.1088/1748-0221/5/03/T03009}}.

\bibitem{bib-crystalball}
\href {http://www.slac.stanford.edu/cgi-wrap/getdoc/slac-r-255.pdf} {J.~E.
  Gaiser, ``Charmonium Spectroscopy from Radiative Decays of the J/$\psi$ and
  $\psi'$''}.
\newblock PhD thesis, SLAC, 1982.
\newblock SLAC-R-255, Appendix F.

\bibitem{bib-pdg}
\hrefCMSnoop {} {{ Particle Data Group} Collaboration, ``2010 Review of
  Particle Physics'',} \textit{ J. Phys. G} \textbf{ 37} (2010) 075021.
  \href{http://dx.doi.org/10.1088/0954-3899/37/7A/075021}{\texttt{
  doi:10.1088/0954-3899/37/7A/075021}}.

\bibitem{bib-faccioli}
\hrefCMSnoop {} {P.~Faccioli {et~al.}, ``Towards the experimental clarification
  of quarkonium polarization'',} \textit{ Eur. Phys. J. C} \textbf{ 69} (2010)
  657. \href{http://dx.doi.org/10.1140/epjc/s10052-010-1420-5}{\texttt{
  doi:10.1140/epjc/s10052-010-1420-5}}.

\bibitem{bib-trackermomentum}
\href {http://cdsweb.cern.ch/record/1279137} {{ CMS} Collaboration,
  ``Measurement of Momentum Scale and Resolution using Low-mass Resonances and
  Cosmic Ray Muons'',} CMS Physics Analysis Summary CMS-PAS-TRK-10-004, (2010).

\bibitem{bib-muonreco}
\href {http://cdsweb.cern.ch/record/1279140} {{ CMS} Collaboration,
  ``Performance of muon identification in pp collisions at $\sqrt{s}$ = 7
  {TeV}'',} CMS Physics Analysis Summary CMS-PAS-MUO-10-002, (2010).

\bibitem{bib-trackingefficiency}
\href {http://cdsweb.cern.ch/record/1279139} {{ CMS} Collaboration,
  ``Measurement of Tracking Efficiency'',} CMS Physics Analysis Summary
  CMS-PAS-TRK-10-002, (2010).

\bibitem{punzibias}
\href {http://www.slac.stanford.edu/econf/C030908/papers/WELT002.pdf}
  {G.~Punzi, ``Comments on likelihood fits with variable resolution'',} in
  \textit{ Proceedings of the PHYSTAT 2003 conference}.
\newblock 2003.
\newblock
\href{http://www.arXiv.org/abs/physics/0401045}{\texttt{
  arXiv:physics/0401045}}.
\newblock

\bibitem{thetables}
\href {http://twiki.cern.ch/twiki/bin/view/CMSPublic/PhysicsResultsBPH10014}
  {{CMS Collaboration}}.
\newblock Numerical results for this paper:
  \url{http://twiki.cern.ch/twiki/bin/view/CMSPublic/PhysicsResultsBPH10014}.

\bibitem{facciolipriv}
\hrefCMSnoop {} {P.~Faccioli, C.~Louren{\c{c}}o, J.~Seixas{ et~al.},
  ``Determination of $\chi_c$ and $\chi_b$ polarizations from dilepton angular
  distributions in radiative decays'',} \textit{ Phys. Rev. D} \textbf{ 83}
  (2011) 096001. \href{http://dx.doi.org/10.1103/PhysRevD.83.096001}{\texttt{
  doi:10.1103/PhysRevD.83.096001}}.

\bibitem{cdfchic1}
\hrefCMSnoop {} {{ CDF} Collaboration, ``Production of \Jpsi\ mesons from
  $\chi_c$ meson decays in $p{\bar p}$ collisions at $\sqrt{s}$ = 1.8 TeV'',}
  \textit{ Phys. Rev. Lett.} \textbf{ 79} (1997) 578.
  \href{http://dx.doi.org/10.1103/PhysRevLett.79.578}{\texttt{
  doi:10.1103/PhysRevLett.79.578}}.

\bibitem{cdfchic2}
\hrefCMSnoop {} {{ CDF} Collaboration, ``Measurement of $\sigma_{\chi_{c2}}
  {\cal B}(\chi_{c2} \to \Jpsi \gamma)/ \sigma_{\chi_{c1}} {\cal B}(\chi_{c1}
  \to \Jpsi\gamma)$ in $p{\bar p}$ Collisions at $\sqrt{s}$ = 1.96 TeV'',}
  \textit{ Phys. Rev. Lett.} \textbf{ 98} (2007) 232001.
  \href{http://dx.doi.org/10.1103/PhysRevLett.98.232001}{\texttt{
  doi:10.1103/PhysRevLett.98.232001}}.

\end{thebibliography}\endgroup

\clearpage
\cleardoublepage \appendix\section{The CMS Collaboration \label{app:collab}}\begin{sloppypar}\hyphenpenalty=5000\widowpenalty=500\clubpenalty=5000\textbf{Yerevan Physics Institute,  Yerevan,  Armenia}\\*[0pt]
S.~Chatrchyan, V.~Khachatryan, A.M.~Sirunyan, A.~Tumasyan
\vskip\cmsinstskip
\textbf{Institut f\"{u}r Hochenergiephysik der OeAW,  Wien,  Austria}\\*[0pt]
W.~Adam, T.~Bergauer, M.~Dragicevic, J.~Er\"{o}, C.~Fabjan, M.~Friedl, R.~Fr\"{u}hwirth, V.M.~Ghete, J.~Hammer\cmsAuthorMark{1}, M.~Hoch, N.~H\"{o}rmann, J.~Hrubec, M.~Jeitler, W.~Kiesenhofer, M.~Krammer, D.~Liko, I.~Mikulec, M.~Pernicka, B.~Rahbaran, H.~Rohringer, R.~Sch\"{o}fbeck, J.~Strauss, A.~Taurok, F.~Teischinger, C.~Trauner, P.~Wagner, W.~Waltenberger, G.~Walzel, E.~Widl, C.-E.~Wulz
% ...
\vskip\cmsinstskip
% ...
1:~~Also at CERN, European Organization for Nuclear Research, Geneva, Switzerland\\
% ...

\end{sloppypar}
\end{document}